\documentclass[manuscript]{aastex}
\usepackage[normalem]{ulem}

\slugcomment{To appear in ApJ}

\shorttitle{Rotationally-Driven Fragmentation for the Formation of L1551\,IRS\,5}
\shortauthors{Lim et al.}

\begin{document}


\title{Rotationally-Driven Fragmentation for the Formation of the Binary Protostellar System L1551\,IRS\,5}


\author{Jeremy Lim}
\affil{Department of Physics, The University of Hong Kong, Pokfulam Road, Hong Kong \\
\&
\\
Laboratory for Space Research, Faculty of Science, The University of Hong Kong, Pokfulam Road, Hong Kong}
\email{jjlim@hku.hk}

\author{Paul K. H. Yeung}
\affil{Department of Physics, The University of Hong Kong, Pokfulam Road, Hong Kong}

\author{Tomoyuki Hanawa}
\affil{Center for Frontier Science, Chiba University, Inage-ku, Chiba 263-8522, Japan}

\author{Shigehisa Takakuwa\altaffilmark{1}}
\affil{Institute of Astronomy and Astrophysics, Academia Sinica, Taipei 10617, Taiwan}

\author{Tomoaki Matsumoto}
\affil{Faculty of Humanity and Environment, Hosei University, Chiyoda-ku, Tokyo 102-8160, Japan}

\and

\author{Kazuya Saigo}
\affil{Department of Physical Science, Graduate School of Science, Osaka Prefecture University, 1-1 Gakuen-cho, Naka-ku, Sakai, Osaka 599-8531, Japan}

\altaffiltext{1}{Current address: Department of Physics and Astronomy, Graduate School of Science and Engineering, Kagoshima University, 1-21-35 Korimoto, Kagoshima, Kagoshima, 890-0065, Japan}



\begin{abstract}
Either bulk rotation or local turbulence is widely invoked to drive fragmentation in collapsing cores so as to produce multiple star systems.  Even when the two mechanisms predict different manners in which the stellar spins and orbits are aligned, subsequent internal or external interactions can drive multiple systems towards or away from alignment thus masking their formation process.  Here, we demonstrate that the geometrical and dynamical relationship between the binary system and its surrounding bulk envelope provide the crucial distinction between fragmentation models.  We find that the circumstellar disks of the binary protostellar system L1551\,IRS\,5 are closely parallel not just with each other but also with their surrounding flattened envelope.  Measurements of the relative proper motion of the binary components spanning nearly 30\,yr indicate an orbital motion in the same sense as the envelope rotation.  Eliminating orbital solutions whereby the circumstellar disks would be tidally truncated to sizes smaller than are observed, the remaining solutions favor a circular or low-eccentricity orbit tilted by up to $\sim$25\degr\ from the circumstellar disks.  Turbulence-driven fragmentation can generate local angular momentum to produce a coplanar binary system, but which bears no particular relationship with its surrounding envelope.  Instead, the observed properties conform with predictions for rotationally-driven fragmentation.  If the fragments were produced at different heights or on opposite sides of the midplane in the flattened central region of a rotating core, the resulting protostars would then exhibit circumstellar disks parallel with the surrounding envelope but tilted from the orbital plane as is observed.
\end{abstract}





\section{Introduction}\label{Introduction}
While there is a generally accepted framework for how single stars form, the manner in which binary and high-order multiple stars form remains contentious.  Yet, the formation of such systems, rather than of single stars, constitutes the primary mode by which the majority of stars having masses comparable to and higher than that of the Sun form.  At the present time, the debate on how binary (hereafter, used generically to also include higher-order multiple) stars form are focussed on two primary issues: (i) what mechanism(s) create the seeds for a binary system, and when are these seeds produced during the collapse of a gravitationally-bound condensation (core); and (ii) what factors determine the growth of the individual protostellar components to reach their final, often different, masses.  Knowledge of the ingredients involved in defining the individual masses of binary components is critical for addressing another outstanding question in star formation, the origin of the stellar initial mass function.

Historically, a number of pathways have been proposed for the formation of binary stars: fission, capture, and fragmentation \citep[e.g., review by][]{Bodenheimer2000, Tohline2002}.  Numerical simulations show that rotational instabilities in self-gravitating spheroids do not lead to fission -- splitting into two (equal) pieces.  Instead, the spheroid develops a central bar and spiral arms, the latter of which transport both angular momentum and mass outwards to stabilize the structure against fission \citep[e.g.,][]{Durisen1986, Williams1988}.  Capture as a result of a close encounter between single protostars is energetically unfavourable, even when large and massive circumstellar disks are invoked to help absorb the impact energy \citep{Clarke1991, Clarke1993}.  Encounters typically result in flybys, with each flyby partially disrupting the disk and reducing its ability to absorb the impact energy of the next encounter.  Recently, gravitational instabilities in circumstellar disks have been invoked to form relatively low-mass binary companions \citep[e.g.,][]{Adams1989}.  Given the large range in mass ratios exhibited by binary systems, however, this process can constitute, at best, a minor pathway for the formation of binary stars.

In part therefore by a process of elimination, fragmentation -- the internal break-up of a core into two or more pieces -- has emerged as the leading contender for how the majority of binary stars form \citep[e.g., review by][]{Goodwin2007}.  Two different mechanisms have been proposed to drive fragmentation: (i) bulk (large-scale ordered) rotation; and (ii) {local (small-scale)} turbulence.  As described in more detail in $\S\ref{Fragmentation Models}$, depending on the actual circumstances involved, these two mechanisms can predict very different geometries and dynamics for the resulting binary system: i.e., alignment between circumstellar disks and/or spin axes of the binary components, as well as alignment between circumstellar disks and orbital plane or between the spin and orbital axes.  Comparisons between binary properties and model predictions for their formation, however, are complicated by possible internal or external interactions during or after the protostellar phase.  Depending on the nature of the interaction, the binary system can be driven either towards or away from alignment, altering its original geometry and dynamics {thus masking its formation process}.  

As we shall demonstrate in this manuscript, a more promising approach to distinguish between different fragmentation models is to study binary systems in the process of formation.  Still embedded in their parental cores, the different proposed drivers of fragmentation make very different predictions for the geometrical and dynamical relationship between the {resulting} protostellar binaries and their surrounding envelopes (see $\S\ref{Fragmentation Models}$).  In this manuscript, incorporating new data having a significantly improved angular resolution and sensitivity, we accurately deduce the alignment between the circumstellar disks, as well as between the circumstellar disks and surrounding envelope, of a binary protostellar system, the Class\,I object L1551\,IRS\,5.  Furthermore, we thoroughly explore best-fit orbits to measurements of relative proper motion for the binary protostars, thus deducing the alignment between the orbital plane and the circumstellar disks.  From the geometrical and dynamical relationship between the binary system and its parental core, 
we reason that the physical properties of L1551\,IRS\,5 reflect the manner in which it formed rather than being imposed by subsequent interactions.  As a consequence, its physical properties can be directly used to differentiate between different mechanisms that drove the fragmentation of its parental core.  The work reported in this manuscript lays the foundation for theoretical simulations on how the binary components in L1551\,IRS\,5 are interacting with and accreting from their envelope, motivating planned observations of this system with the Atacama Large Millimeter and Submillimeter Array (ALMA).  We recently conducted such a theoretical simulation for the Class\,I object L1551\,NE, assuming a circular coplanar orbit for the binary protostars as well as a coplanar circumbinary disk.  We found the model predictions to be in good agreement with the observed structure and kinematics of its circumbinary disk as measured with ALMA \citep{Takakuwa2014}.

This manuscript is arranged as follows.  In $\S\ref{Observations}$, we describe our most recent observation of the circumstellar disks of the binary protostars in L1551\,IRS\,5, surpassing in both sensitivity and angular resolution previous observations.  The results are presented in $\S\ref{Results}$, along with an explanation of how we separated the ionized jets from the circumstellar dust disks.  In $\S\ref{Physical Parameters Circumstellar Disks}$, we describe how we determine the inclinations, relative alignments, and sizes of the binary circumstellar disks.  In $\S\ref{Physical Parameters Envelope}$, we describe how we determine the geometry of their surrounding envelope, how this geometry relates to that of the circumstellar disks, and place constraints on a central cavity in the envelope.  In $\S\ref{Orbit}$, we present all currently available measurements for the relative proper motion of the binary protostars, acceptable best-fit orbits to these measurements, and constraints placed on the orbits by the sizes of the circumstellar disks and tentative upper limits on the size of any central cavity.  In $\S\ref{Discussion}$, we bring together all the aforementioned properties of L1551\,IRS5 to assemble a coherent picture for how this system formed.  We also extrapolate into the future the likely evolution of L1551\,IRS\,5, and how its projected properties on the main sequence will compare with those of other binary systems having similar masses and orbital separations.  In $\S\ref{Summary}$, we provide a thorough summary of our results, analyses, and interpretation. Throughout this manuscript, we assume a distance to L1551\,IRS\,5 of 140\,pc.

\section{Observations and Data Reduction}\label{Observations}
We observed L1551\,IRS\,5 using the Jansky Very Large Array (VLA) on 2012 November 16, 28, and 29, spanning a duration of $\sim$2.5\,hr on each day.  The correlator was configured to provide a bandwidth of 8\,GHz over the frequency range 39--47\,GHz (central wavelength of $\sim$7\,mm).  To mitigate against rapid changes in absorption and refraction by the Earth's atmosphere, causing rapid fluctuations in the measured visibility amplitude and phase of the target source, we switched between observations of L1551\,IRS\,5 and a nearby secondary calibrator, the quasar J0431$+$1731, every 20\,s.  As a check of the quality of the amplitude and phase corrections, we performed similar observations of a quasar lying close to L1551\,IRS\,5, J0431$+$2037, every $\sim$30\,mins.  This quasar also was used to check the pointing accuracy of the antennas, a task performed every $\sim$1\,hr.  The bright quasar J0510+1800 served as the bandpass calibrator, and the quasar 3C48 as the flux calibrator.

We edited, calibrated, and made maps from the data using the Common Astronomy Software Applications (CASA) package.  Considerable effort went into weeding out radio-frequency interference (RFI), which can be very weak and difficult to find, to ensure that the actual data used for making the maps is as free of contamination as is possible.  The calibration was performed in the standard manner (e.g., examples in https://casaguides.nrao.edu/index.php/Karl\_G.\_Jansky\_VLA\_Tutorials) as recommended by the observatory.  Maps were made using two different weighting schemes, natural (i.e., equal weights on all visibilities) and $\rm Robust = -1.0$ (unequal weights designed to provide a more uniform sampling in $uv$-space), to accentuate different features of interest.  The synthesized beams and root-mean-square (rms) noise fluctuations ($\sigma$) of the maps thus made are summarised in Table\,\ref{Map Parameters}.  Notice that the synthesized beams obtained using the two weighting schemes are close to circular, making it easier to visually interpret as well as analyze the maps.   All subsequent analyses of the images obtained were made using the Astronomical Image Processing System (AIPS) and GALFIT software packages.

\section{Results and Analyses}\label{Results}

\subsection{Images}\label{Images}
Figure\,\ref{L1551IRS5 image natural}($a$) (color map, along with contours plotted at varying intervals of $\sigma$) and Figure\,\ref{L1551IRS5 image natural}($d$) (same color map, along with contours now plotted at uniformly spaced percentage levels of the peak intensity) show the map obtained for L1551\,IRS\,5 using natural weighting, yielding an angular resolution of $55.6 {\rm \ mas} \times 52.7 {\rm \ mas}$ ($7.8 {\rm \ AU} \times 7.4 {\rm \ AU}$).  Two sources are apparent, separated by $0\farcs36$ and aligned almost exactly along the north-south direction.  
We shall henceforth refer to the more northerly source as the N source, and the more southerly source as the S source.  Both sources are clearly elongated along the north-west to south-east directions.  In addition, relatively weak protrusions can been seen extending away from each source towards the north-east and south-west, oriented in approximately the same directions as a pair of ionized jets imaged at 3.5\,cm with the VLA by \citet{Rodriguez2003b}.  \citet{Rodriguez2003b} infer a position angle of $67\degr \pm 3\degr$ for the northern and $55\degr \pm 1\degr$ for the southern jet, as indicated by the arrows in Figure\,\ref{L1551IRS5 image natural}$a$.  Notice that the protrusion on the south-western side of the S source lies northwards of the inferred jet axis for this component.  The same offset also is apparent for this part of the jet in its image at 3.5\,cm as pointed out by \citet{Rodriguez2003b}.  Finally, even weaker protrusions can be seen on the northern and south-eastern sides of the S source, and perhaps (at close to marginal significance) also on the northern and southern sides of the N source.

Figure\,\ref{L1551IRS5 image robust}($a$) (color map, along with contours plotted at varying intervals of $\sigma$) and Figure\,\ref{L1551IRS5 image robust}($d$) (same color map, along with contours now plotted at uniformly spaced percentage levels of the peak intensity) show the map obtained using robust weighting, providing nearly a factor of two higher angular resolution of $33.4 {\rm \ mas} \times 31.7  {\rm \ mas}$ ($4.7 {\rm \ AU} \times 4.4 {\rm \ AU}$) but a factor of about three poorer sensitivity.  
Apart from the strongest protrusion on the north-eastern side of the S source (related to its jet), all the other protrusions seen in the naturally-weighted map are not detectable in the robust-weighted map.  As before, both sources are clearly elongated along the north-west to south-east directions.  At or close to their centers, both sources appear to be elongated also along the north-east to south-west directions, quite closely aligned with the inferred orientations of their ionized jets.

In our previous observation of L1551\,IRS\,5 at 7\,mm with the VLA (before its upgrade), we attained a far inferior sensitivity, along with a somewhat poorer angular resolution despite utilizing the Pie Town antenna as part of the array \citep{Lim2006}.  In that observation, we detected a relatively compact feature at a 5$\sigma$ significance level southeast of the N source.  In the observation reported here, apart from the N and S sources, we detect no other source in our field of view (extending far beyond the binary system to a full-width half-maximum of 1\arcmin).  Although we cannot rule out a transient source, it seems more likely that the weak and relatively compact feature detected in our previous observation is an artefact produced by the poorer quality of the calibration possible at the time (compared with the case here where, because of the much higher instantaneous sensitivity, corrections for variations in atmospheric phase can be made on much shorter timescales and at a much higher precision).  

\subsection{Decomposing Circumstellar Dust Disks and Ionized Jets}\label{Decomposing Circumstellar Disks and Jets}\label{Decomposing}
\begin{figure}
\center
\vspace{-8cm}
\epsscale{1.2}
\plotone{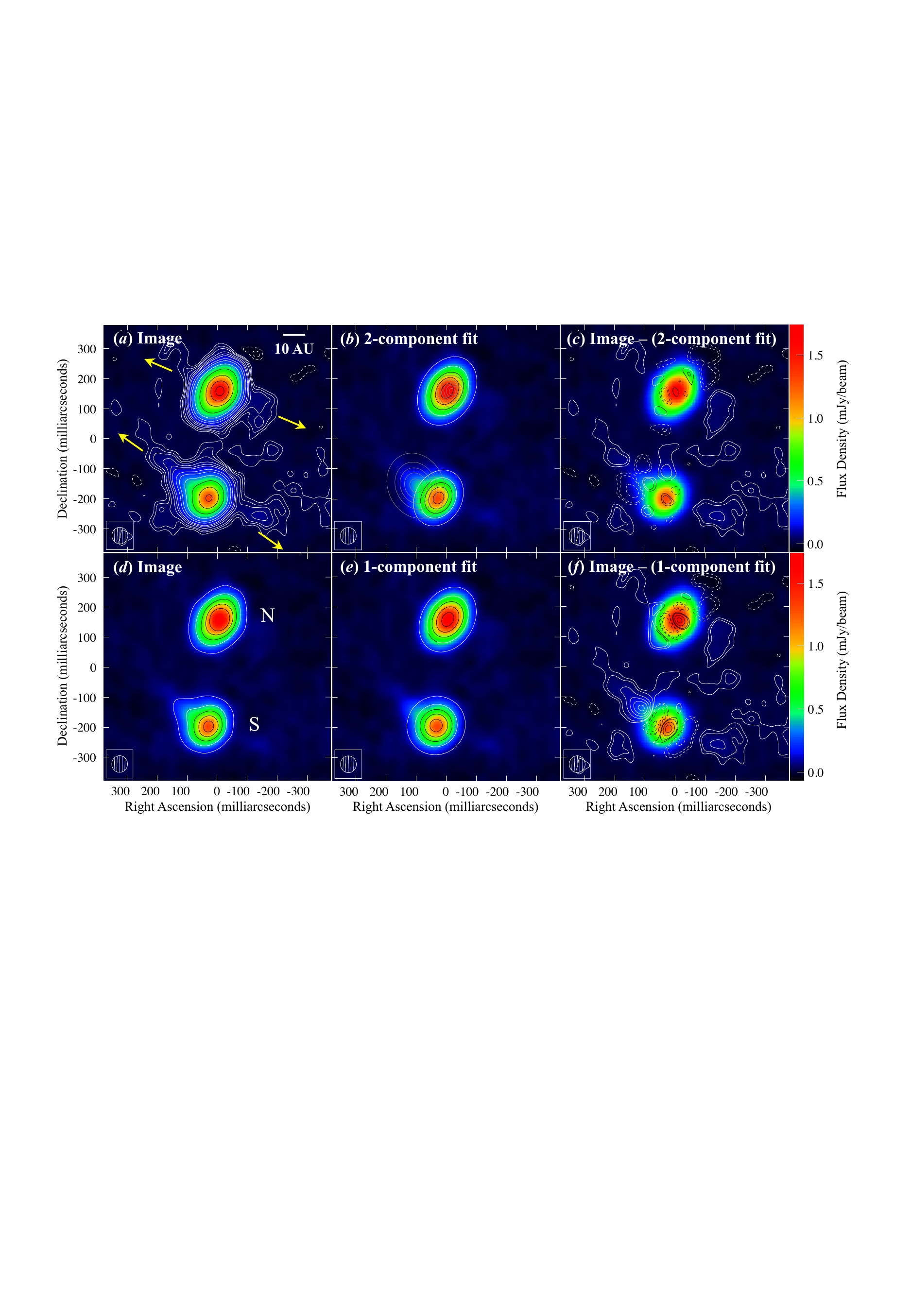}
\vspace{-11.5cm}
\renewcommand{\baselinestretch}{1.3}
\caption{Color map (all panels) of L1551\,IRS\,5 made at 7\,mm with the VLA using natural weighting.  The more northerly source is referred to as the N source, and the more southerly as the S source.  The synthesized beam is shown at the lower left corner of each panel.  $(a)$ Contours of the color map with levels plotted at (-3, -2, 2, 3, 4, 6, 8, 10, 15, 20, 30, 50, 80, 110, \& $140) \times \sigma$, where $\sigma = 1.15 \times 10^{-5} {\rm \ Jy \ beam^{-1}}$ (rms noise level).  Arrows indicate orientations of ionized jets at 3.6\,cm inferred by \citet{Rodriguez2003b}.  $(b)$ Contours of two-component, two-dimensional, Gaussian function fitted to each source, plotted at $10\%$, $30\%$, ... $90\%$ of the peak intensity of the individual components to permit a simple visual comparison of their relative sizes.  $(c)$ Residuals after subtracting the fitted Gaussian components in panel $(b)$ from the color map, revealing that these components closely capture the main bodies of the two sources.  $(d)$ Contours of the color map with levels plotted at $10\%$, $30\%$, ... $90\%$ of the peak intensity of each source.  $(e)$ Contours of a single two-dimensional Gaussian function fitted to each source, plotted at $10\%$, $30\%$, ... $90\%$ of the peak intensity of each function.  Comparison of the contour plots in panels $(d)$ and $(e)$ provides a simple visual impression of how well a single Gaussian component fits the main body of each source.  $(f)$ Residuals after subtracting the fitted Gaussian components in panel $(e)$ from the color map, revealing that a single Gaussian component poorly captures the main body of either source.  The strong residuals in the main bodies correspond to one of the two Gaussian components fitted to the respective sources in panel $(b)$.}
\label{L1551IRS5 image natural}
\end{figure}

\begin{figure}
\center
\epsscale{1.2}
\plotone{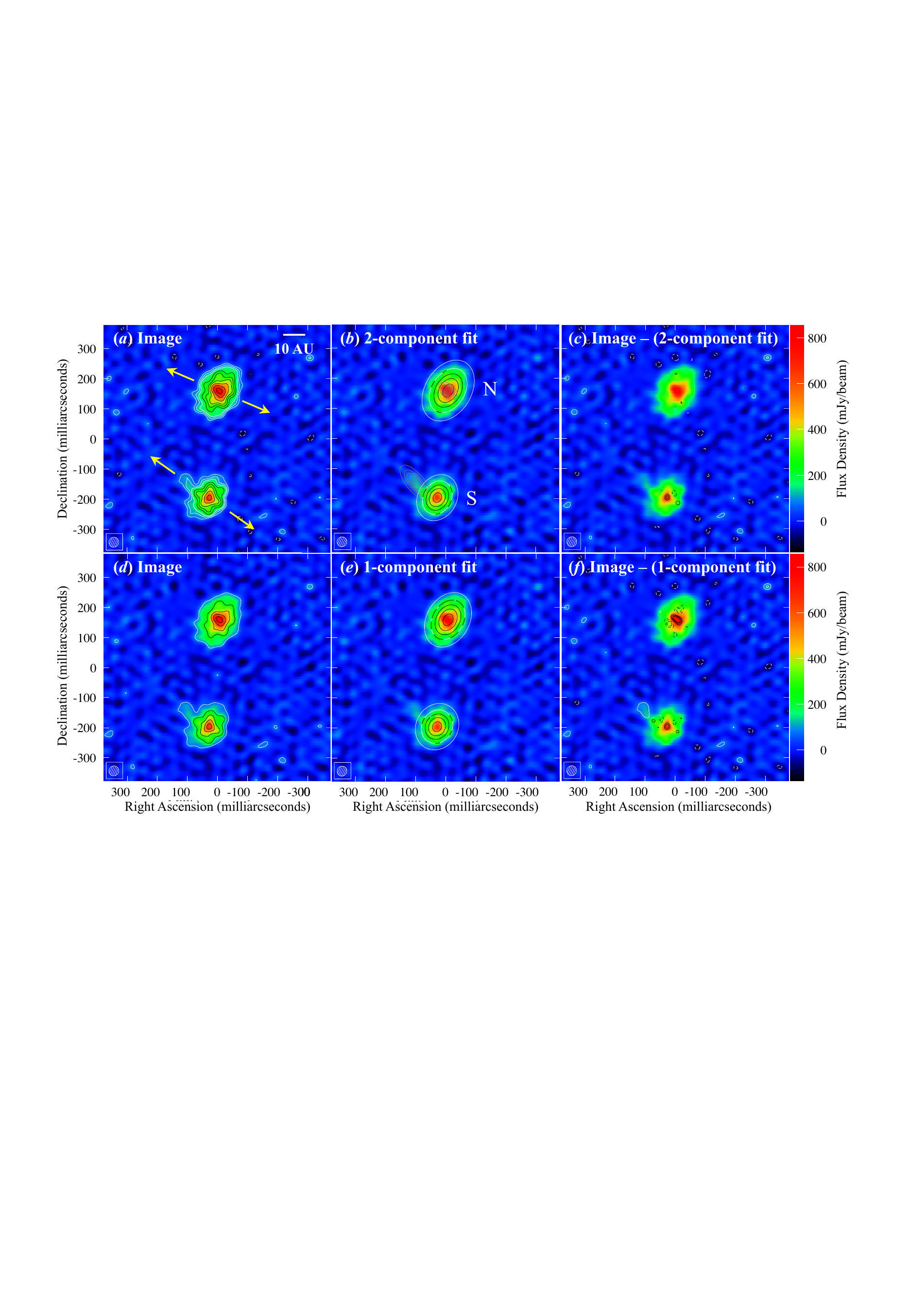}
\vspace{-11cm}
\renewcommand{\baselinestretch}{1.3}
\caption{Same as for Fig.\,\ref{L1551IRS5 image natural}, but now with $\rm Robust = -1.0$ weighting of the visibilities to provide a higher angular resolution albeit poorer sensitivity.}
\label{L1551IRS5 image robust}
\end{figure}

As pointed out by \citet{Rodriguez1998}, at 7\,mm the emission from the binary components in L1551\,IRS\,5 is dominated by dust from their circumstellar disks.  When observed at a sufficiently high sensitivity, however, free-free emission from their ionized jets also is detectable at 7 mm, as found by \citet{Lim2006} and as is clearly apparent in Figure\,\ref{L1551IRS5 image natural}($a$) and to a lesser extent also Figure\,\ref{L1551IRS5 image robust}($a$).  
To determine the geometry of the circumstellar disks, one of the main goals of this work, we therefore have to remove the contributions from the ionized jets.  

As can be seen in, especially, the naturally-weighted map of Figure\,\ref{L1551IRS5 image natural}($a$), 
the ionized jets from both protostars have a complicated structure.  Fortunately, to separate the ionized jets from the circumstellar disks, we only have to subtract the portion of the jet projected against each circumstellar disk rather than the entire jet structure.  As we shall show, a two-component 2-dimensional Gaussian function fitted simultaneously to each source provides an adequate decomposition of
the projected portions of the jets from the respective circumstellar disks.  The parameters of the dual Gaussian components that provide the best fit to each source in the natural- and robust-weighted maps separately are listed in Table\,\ref{Fitting Parameters}.  These components, convolved with the corresponding synthesized beams (also two-dimensional Gaussian functions with parameters as listed in Table\,\ref{Map Parameters}), are plotted in contours in Figures\,\ref{L1551IRS5 image natural}($b$) and \ref{L1551IRS5 image robust}($b$), overlaid on the color maps of Figures\,\ref{L1551IRS5 image natural}($a$) and \ref{L1551IRS5 image robust}($a$) respectively.  To provide a simple visual comparison of the relative sizes of the Gaussian components fitted to each source, the contour levels for each Gaussian component are plotted at the same percentage levels of the peak intensity of that component.  As can be seen, for each source, one of the fitted Gaussian components has its major axis closely aligned with the inferred orientation of the jet from that source (indicated by the arrows in Figures\,\ref{L1551IRS5 image natural}($a$) and \ref{L1551IRS5 image robust}($a$)), and must therefore correspond to the inner portion of the jet.  The other component has its major axis closely orthogonal to the inferred jet orientation, and must therefore correspond to the circumstellar disk.  

The residuals after subtracting the two-component, two-dimensional Gaussian components fitted to each source are shown in contours in Figure\,\ref{L1551IRS5 image natural}($c$) for the natural- and Figure\,\ref{L1551IRS5 image robust}($c$) for the robust-weighted map, again overlaid on the color maps of Figures\,\ref{L1551IRS5 image natural}($a$) and \ref{L1551IRS5 image robust}($a$) respectively.  Although significant residuals ($\leq$$-3\sigma$ and/or $\geq$$3\sigma$) are apparent in Figure\,\ref{L1551IRS5 image natural}($c$), these residuals coincide primarily with protrusions outside the main bodies --- and therefore also circumstellar disks --- of the N and S sources.  Within the main bodies of the two sources, all the residuals are $< 4\sigma$.  Thus, in the naturally-weighted maps, the fitted Gaussian components provide a reasonably good decomposition of the projected portions of the jets from the circumstellar disks.  In Figure\,\ref{L1551IRS5 image robust}($c$), the only significant residual ($3.5\sigma$) is that at the centroid of the S source.  
Thus, in the uniformly-weighted maps, the fitted Gaussian components provide an excellent decomposition of the projected portions of the jets from the circumstellar disks.  Subtracting the Gaussian components fitted to the jets leave the jet-subtracted images of the N and S sources shown in Figure\,\ref{NUKER} (first column), used in $\S\ref{Physical Parameters Circumstellar Disks}$ for determining the physical parameters of the circumstellar disks.

As a simple demonstration of the need to subtract the jets so as to obtain a good fit for the circumstellar disks, in Figures\,\ref{L1551IRS5 image natural}($e$) and \ref{L1551IRS5 image robust}($e$) we plot in contours just a single 2-dimensional Gaussian function fitted to each of the two sources.
Once again, these Gaussian components are plotted at the same percentage levels of their individual peak intensities.  To provide a simple visual comparison of how well these components actually match the main bodies of the N and S sources, in Figures\,\,\ref{L1551IRS5 image natural}($d$) and \ref{L1551IRS5 image robust}($d$) we plot contour levels for the two sources at the same percentage levels of their individual peak intensities.  As can be seen, there are clear differences between the fitted and observed structures in the central region of the N source as well as the north-eastern side of the S source.  These differences correspond to the projected portions of the jets captured in the two-component fit to each source as described above.  The same differences appear as relatively strong residuals in Figures\,\ref{L1551IRS5 image natural}($f$) and \ref{L1551IRS5 image robust}($f$) after subtracting the single Gaussian component fitted to each source.  


As listed in Table\,\ref{Fitting Parameters}, both the peak and integrated intensities of the Gaussian components fitted to the circumstellar disks are higher (by up to an order of magnitude) than those fitted to their corresponding ionized jets.  Thus, at 7\,mm, emission from the circumstellar disks dominate over that from the jets projected closely against these disks, consistent with that found previously.  Furthermore, the centroids of the Gaussian components fitted to the circumstellar disks coincide with their respective source centroids.  By contrast, the centroids of the Gaussian components fitted to the jets are clearly displaced from their respective source centroids in the natural- and, for the S source, also the robust-weighted maps.  Finally, the circumstellar disks are resolved along their minor as well as major axes in both the natural- and robust-weighted maps.  By contrast, only the southern jet is resolved along both its major and minor axes, and then only in the natural-weighted map.

\section{Physical Parameters of Circumstellar Disks}\label{Physical Parameters Circumstellar Disks}

At the spatial resolutions so far attained in observations of protostars, it is not possible to uniquely constrain the radial intensity profiles of their circumstellar disks.  
For example, as shown in $\S\ref{Decomposing}$, a two-dimensional Gaussian function provides a good fit to the image of each of the two circumstellar disks in L1551\,IRS5.  
Circumstellar disks, however, are unlikely to have Gaussian radial intensity profiles.  Instead, images of the circumstellar dust disks of protostars are usually fitted with physically motivated models for their radial intensity profiles.

For single protostars, a power-law radial surface brightness distribution is commonly assumed for their circumstellar dust disks.  This brightness profile is designed to mimic disks that have power-law surface density and temperature profiles \citep[e.g., see brief review by][]{Hughes2008}.  In such models, the disk is usually truncated at an inner radius that reflects either its innermost physical extent or that within which dust sublimates (but gas is present), as well as at an outer radius that reflects its outermost physical extent.  The outer disk radius derived in such models depends on the value inferred or chosen for its inner radius.  In pre-main-sequence stars, the inner radius of the dust disk can be inferred by modelling its spectral energy distribution at near- to mid-infrared wavelengths.  This approach, however, is not possible for protostars, which are obscured by their optically-thick envelopes at near- to mid-infrared wavelengths.  To further complicate matters, models that invoke a sharply truncated outer edge have been found to yield an outer radius for the dust component significantly smaller than that seen in scattered light or the outer radius for the gaseous (as traced in CO) component \citep[][and references therein]{Hughes2008}.  Instead, \citet{Hughes2008} find that models in which the disk tapers off gradually at its edge can better reproduce, simultaneously, the emission observed from both the dust and gas components.

\begin{figure}
\epsscale{1.2}
{\hspace{-2cm}\plotone{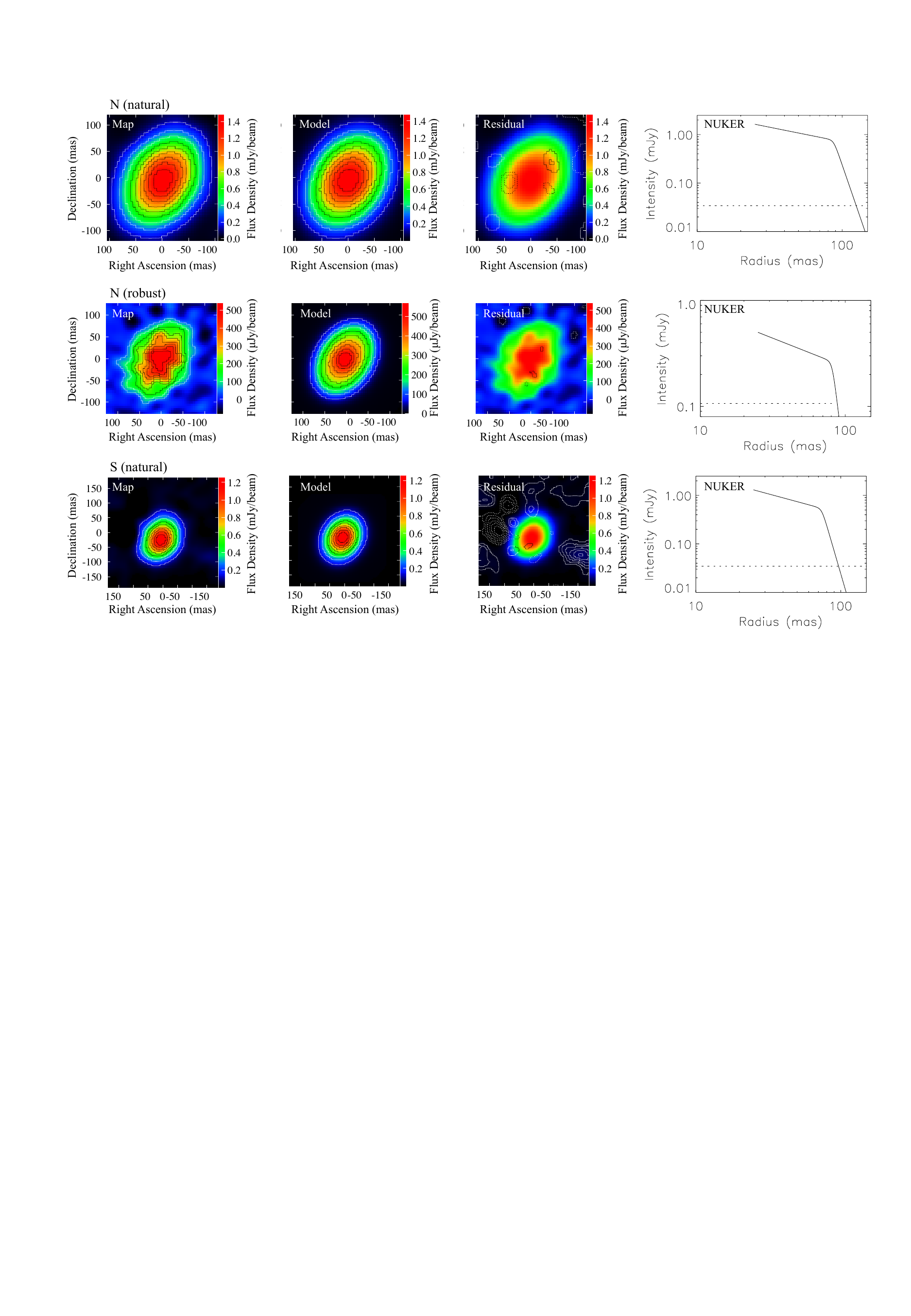}}
\vspace{-16cm}
\renewcommand{\baselinestretch}{1.3}
\caption{Jet-subtracted maps (first column), NUKER models (second column) fitted to each source with parameters as listed in Table\,\ref{Parameters NUKER fits}, residuals plotted in contours after subtracting the NUKER models from the jet-subtracted maps plotted in color (third column), and the profile of the fitted NUKER model along the major axis (last column).  Results for the natural-weighted map of the N source (upper row), robust-weighted map of the N source (middle row), and naturally-weighted map of the S source (lower row), are presented separately.  Contours are plotted at $10\%$, $30\%$, ... $90\%$ of the peak intensity in the jet-subtracted maps and NUKER models.  Contours are plotted at -3, -2, 2, $3 \times \sigma$, where $\sigma$ is the rms noise level (Table\,\ref{Map Parameters}), in the residual map for the N source, and -7, -6, ..., -2, 2, 3, ... $7 \times \sigma$ in the residual map for the S source.}
\label{NUKER}
\end{figure}

By contrast with single protostars, the circumstellar disks of closely-separated binary protostars are predicted to be truncated by tidal interactions with their companions and therefore have sharp edges.  As shown by \citet{Pichardo2005}, tidal forces exerted by one protostar induce elliptical orbits in the circumstellar disk of the other protostar.  Because the orbital eccentricity generally increases with distance from the protostar, outer orbits eventually intersect with inner orbits.  In this situation, the largest non-intersecting orbit defines the outermost stable orbit and hence maximal possible extent of a circumstellar disk.  Motivated by this prediction, we have fitted a two-component power-law radial brightness profile to the circumstellar disks of the N and S sources, where the much steeper outer power-law represents the rapid tapering of the disks at their edges.  To permit a smooth transition between the two power laws (which we deem to be more physical), we fitted a profile known as the NUKER function \citep{Lauer1995}, examples of which are shown in Figure\,\ref{NUKER} (last column).  This function is parameterised as: 
\begin{equation}
I(r) = I_b \ 2^{{\beta-\gamma}\over\alpha} ({r \over r_b})^{-\gamma} [1 + ({r \over r_b})^\alpha]^{\gamma-\beta \over \alpha} \ \ \ , 
\end{equation}
where $I(r)$ is the intensity, $I$, as a function of radius, $r$, $\gamma$ is the inner power-law slope, $\beta$ the outer power-law slope, $\alpha$ controls the sharpness of the transition between the two power laws (larger $\alpha$ indicating a sharper transition), $r_b$ the break radius at which the slope is the average of $\beta$ and $\gamma$ or, equivalently, the radius of maximum curvature in logarithmic units, and $I_b$ the intensity at $r_b$.  

We fitted two-dimensional NUKER functions to the jet-subtracted images of the N and S sources shown in Figure\,\ref{NUKER} (first column) using the software package GALFIT.  This package was originally developed by \citet{Peng2002}, and its latest incarnation is described by \citet{Peng2010}.  For the natural-weighted map of the N source shown in Figure\,\ref{NUKER} (upper row), we had to tightly restrict the region fitted to an area that just encompasses the main body of this source.  No such restriction was necessary for the naturally-weighted map of the S source shown in Figure\,\ref{NUKER} (middle row) or the robust-weighted map of the N source shown in Figure\,\ref{NUKER} (bottom row), where the jet-related emission outside the main body of each source is relatively weak and do not significant affect the fits.  Sensibly, the GALFIT program does not try to fit for the (spatially unresolved) central region of the circumstellar disk within an area spanned by the FWHM of the synthesized beam.  The results of the fits (the program returns parameters for the lowest $\chi^2$) therefore do not depend on the presence of any central clearing in the circumstellar disk, and makes no statement about the radial brightness profile within this central area.  For completeness, we note that a single power-law (i.e., setting $\beta = \gamma$) does not provide satisfactory fits to either circumstellar disks.

\subsection{Inclination}\label{Inclination Circumstellar Disks}
Leaving all parameters in the NUKER function to be freely fitted, the GALFIT program was not able to find satisfactory fits for either circumstellar disks.  On the other hand, by fixing $\alpha$ (which determines the sharpness of the transition between the inner and outer power-law) at different values, we found that the position angle of the major axis and inclination (as determined from the ratio of the minor to major axes) of both circumstellar disks to be essentially constant independent of $\alpha$ in both the natural- and robust-weighted maps.  Furthermore, these parameters are similar to those derived by fitting a two-dimensional Gaussian function to each circumstellar disk (to within the uncertainties of the Gaussian fits, as listed in Table\,\ref{Fitting Parameters}).
Evidently, although the radial intensity profiles of the circumstellar disks cannot be uniquely constrained, their position angles and inclinations are well determined.  
We therefore fixed these parameters at their formal values derived from the Gaussian fits.

When deriving their inclinations, we assumed implicitly that both circumstellar disks have circularly symmetric radial intensity profiles and are geometrically thin (i.e., vertical thickness much smaller than the disk diameter).  
\citet{Pichardo2005} show that the radius at which orbits in the circumstellar disks significantly depart from near circular only occurs beyond about 80\% of their tidally-truncated radius.  The Gaussian functions fitted to the circumstellar disks have radii at FWHM (the dimemsions used to infer the disk inclinations) that are only about 60\%--70\% of their break radii found by fitting NUKER functions ($\S\ref{Sizes}$).  As a consequence, tidal distortions should not significantly affect the inclinations we infer for the circumstellar disks.  More problematically, if the circumstellar disks are strongly flared, then their inclinations may be underestimated.  Vertical settling of dust to the midplane, if it occurs, can mitigate this problem for disks traced in dust emission, as is the case here.  As we will show in $\S\ref{Orbital Solutions}$, the inferred inclinations of both circumstellar disks are systematically lower than the inclination of the orbital plane for many of the best-fit acceptable orbits, which span a wide range of eccentricities.  The possibility that the circumstellar disks are strongly flared and hence their inclinations underestimated may explain their apparent misalignment with the orbital plane.  On the other hand, as we show in $\S\ref{Inclination Envelope}$, both circumstellar disks are closely parallel with their surrounding flattened envelope, arguing against a large systematic error in their inferred inclinations.

\subsection{Alignment}\label{Alignment}
{ As listed in Table\,\ref{Fitting Parameters}, whether derived from the natural- or robust-weighted maps, the position angles of both circumstellar disks are identical to within the measurement uncertainties.  From the natural-weighted map, the position angle is $148\degr.2 \pm 0\degr.7$ for the circumstellar disk of the N source and $147\degr.7 \pm 1\degr.2$ for the circumstellar disk of the S source, similar to within $0\degr.5 \pm 1\degr.4$.  The uncertainties in the position angles are larger in the robust-weighted maps, where the position angles of the two circumstellar disks are similar to within $2\degr.3 \pm 7\degr.4$.  The inclination of each circumstellar disk as derived from the natural- and robust-weighted maps also is similar to within the measurement uncertainties, although there is a small difference between the inclinations of the two disks in both these maps.  In the naturally-weighted map, the circumstellar disk of the N source has an inclination of $48\degr.1 \pm 0\degr.5$ and that for the S source an inclination of $44\degr.3 \pm 0\degr.8$, a difference of just $3\degr.8 \pm 0\degr.9$.  The uncertainties in the inclinations are larger in the robust-weighted maps, where the inclinations of the two circumstellar disks differ by $9\degr.1 \pm 5\degr.0$.  Henceforth, we shall adopt the position angles and inclinations obtained from the natural-weighted map, where the uncertainties in both these parameters are significantly smaller than those obtained from the robust-weighted maps; this map has a higher sensitivity than the robust-weighted map, and hence samples a larger extent of the circumstellar disks.}  {\it The circumstellar disks of the binary protostars in L1551\,IRS\,5 are therefore very closely parallel}, providing the first of several constraints that we will impose on models for the formation of this binary system.

In binary systems such as L1551\,IRS\,5 where ionized jets can be traced far beyond the circumstellar disks, the alignment between the jets as projected onto the sky plane can potentially serve as a sensitive probe of the alignment between the circumstellar disks.
As mentioned earlier, from a high angular-resolution map at 3.6\,cm where there is no apparent emission from the circumstellar disks, \citet{Rodriguez2003b} infer a position angle of $67\degr \pm 3\degr$ for the northern and $55\degr \pm 1\degr$ for the southern jet.  If the jets emerge perpendicular to their respective circumstellar disks, then the major axes of the two circumstellar disks should have positions angles that differ by $12\degr \pm 3\degr$.  Instead, we find that the major axes of the two circumstellar disks are aligned to within $0\degr.5 \pm 1\degr.4$.

As we will now show, the orientations that \citet{Rodriguez2003b} infer for the two jets may differ from the true orientations at their bases.  At 3.6 cm, both jets in L1551 IRS5 are detected as a series of knots.  The knots are not aligned along a single axis for either the northern or southern jets.  As a consequence, to infer the position angles of the individual jets, \citet{Rodriguez2003b} fitted a two-dimensional Gaussian function to the knot closest to the individual protostellar components \citep[the latter defined by the centroids of their circumstellar disks as measured by][]{Rodriguez1998}.  The knots that \citet{Rodriguez2003b} selected are the same as those fitted by the Gaussian components corresponding to the jet of the N source in Figure\,\ref{L1551IRS5 image natural}$b$ and the jet of the S source in Figure\,\ref{L1551IRS5 image natural}$b$ and \ref{L1551IRS5 image robust}$b$.  As can be seen even in the map at 3.6\,cm presented by \citet{Rodriguez2003b}, the centroid of the knot closest to the S source is visibly displaced to the north-east of this source.  We see the same displacement at 7 mm in both Figures\,\ref{L1551IRS5 image natural}$b$ and \ref{L1551IRS5 image robust}$b$, confirming that this knot does not trace the base of the southern jet.  In the map at 3.6\,cm presented by \citet{Rodriguez2003b}, the centroid of the knot closest to the N source appears to coincide with this source.  At the higher angular resolution of our natural-weighted map at 7 mm (Figure\,\ref{L1551IRS5 image natural}$b$), however, the centroid of this knot can be seen to be displaced to the south-west of the N source.  Thus, this knot does not actually trace the base of the northern jet.  

\begin{figure}
\center
\epsscale{1.3}
\plotone{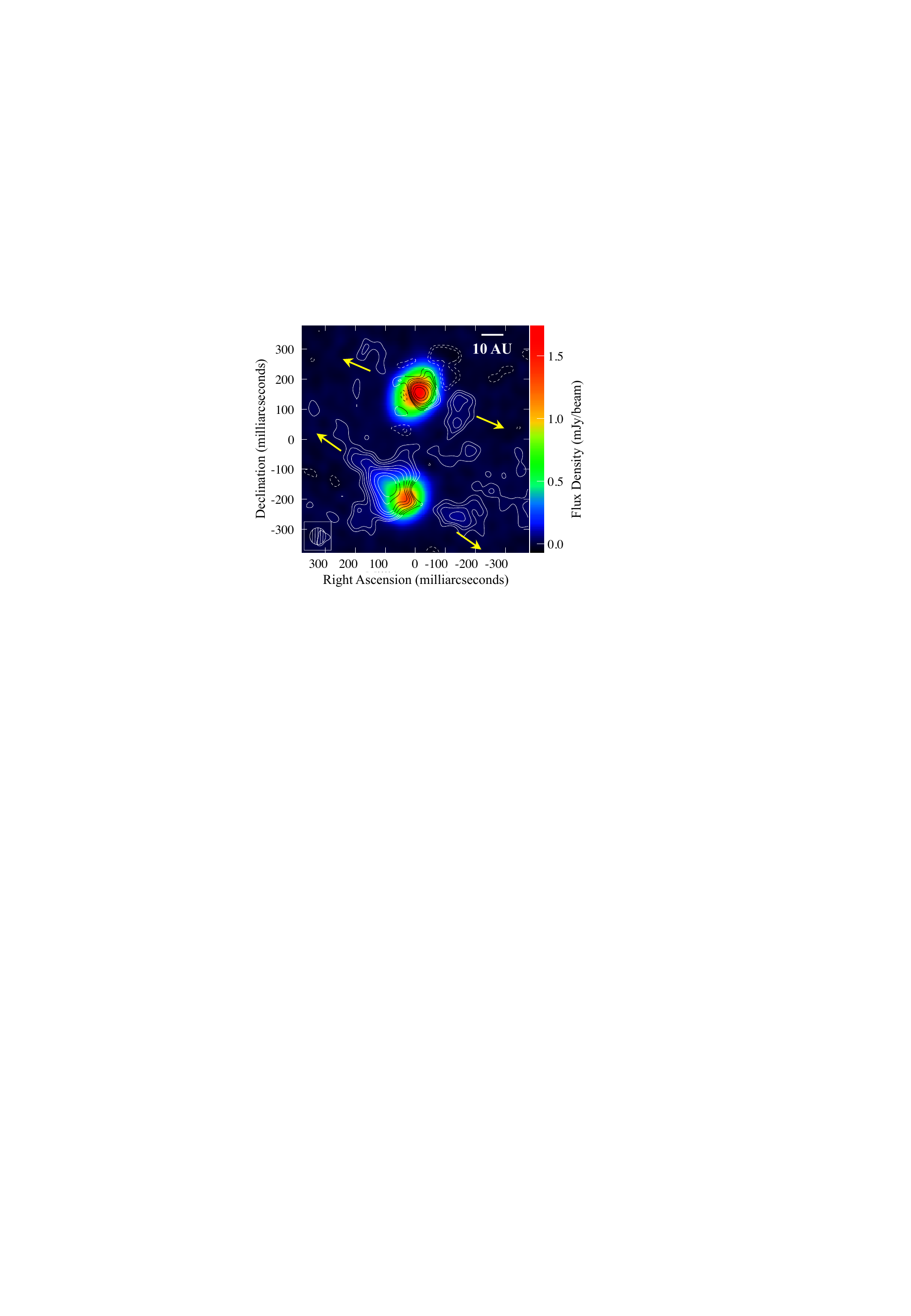}
\vspace{-17.5cm}
\renewcommand{\baselinestretch}{1.3}
\caption{Color map of L1551\,IRS\,5 made at 7\,mm with the VLA using natural weighting (i.e., same as in Fig.\,\ref{L1551IRS5 image natural}).  The synthesized beam is shown at the lower left corner.  Contours show, primarily, emission from the jets after the Gaussian components fitted to the circumstellar disks have been subtracted.  As can be seen, the knot projected against the main body of each source does not coincide with the source centroid, indicating that these knots do not trace the base of the jets.  Furthermore, the knot projected against the main body of the S source is resolved along both its major and minor axes, and exhibits a complex structure.  Caution must therefore be applied when using the major axes of these knots to define the jet axes.  The arrows indicate the position angles of the major axes of the same knots as determined by \citet{Rodriguez2003b} from their map at 3.6\,cm made a decade earlier.}
\label{Ionized Jets}
\end{figure}

From our naturally-weighted map at 7\,mm (Figure\,\ref{L1551IRS5 image natural}$b$), we infer a position angle of $42\degr.3 \pm 8\degr.9$ for the northern and $27\degr.2 \pm 6\degr.2$ for the southern jet (Table\,\ref{Fitting Parameters}) based on the same knots as those used by \citet{Rodriguez2003b}.  These values are significantly different (by $24\degr.7 \pm 9\degr.4$ for the northern and $27\degr.8 \pm 6\degr.3$ for the southern jet) from those determined by \citet{Rodriguez2003b} at a lower angular resolution at 3.6 cm.  Figure\,\ref{Ionized Jets} shows the structure of the knots closest to the N and S sources after the Gaussian components fitted to their circumstellar disks have been subtracted from the natural-weighted map of Figure\,\ref{L1551IRS5 image natural}.  Evidently, the knot closest to the S source has a complicated structure.  Indeed, this knot is resolved not only along its major but also its minor axes; thus, the two-dimensional intensity distribution of this knot affects the position angle inferred for its major axis.  The same also is true for this knot at 3.6\,cm, where it is resolved along both its major and minor axes \citep[see Table\,1 of][]{Rodriguez2003b}.  Thus, in L1551\,IRS\,5, where the knots defining {either of} the two jets do not lie along a single axis and the knot closest to each protostar does not actually trace the base of the jet, we caution against using the apparent misalignment between the jets to infer a corresponding misalignment between their associated circumstellar disks.

\subsection{Sizes}\label{Sizes}

Fixing the position {angles} and {inclinations} of the circumstellar disks to the values listed in Table\,\ref{NUKER} (their formal values as derived from  Gaussian fits {to the N and S sources and listed in Table\,\ref{Fitting Parameters}}), we then varied the value of $\alpha$ to investigate how the remaining free parameters changed with $\alpha$.  For the jet-subtracted image of the N source, we found that the solutions converge to a common set of values for $\gamma$, $\beta$, $r_b$, and $I_b$ as $\alpha$ increases (i.e., sharper transition between the inner and outer power-laws).  For the circumstellar disk of the S source, we found two families of acceptable fits for both the natural- and robust-weighted images.  Only one of the two families in the natural-weighted image, whereas neither families in the robust-weighted image, converges to a common set of values for $\gamma$, $\beta$, $r_b$, and $I_b$ as $\alpha$ increases.  Table\,\ref{Parameters NUKER fits} lists the parameters from the NUKER fits to the jet-subtracted images the N source (both natural- and robust-weighted maps) and S source (natural-weighted image only) at the largest value of $\alpha$ for which a solution exists.  The second column of Figure\,\ref{NUKER} shows {images of} the fitted NUKER {profiles} convolved by the synthesized beam.  The third column shows the residuals, plotted in contours, after subtracting the fitted NUKER profile from the source.   The last column shows the actual intensity profile of the fitted NUKER function along the major axis.

Although the fitted NUKER profiles have different values of $\gamma$ and $\beta$ in the natural- and robust-weighted maps for the circumstellar disk of the N source, the inferred break radius of this disk (where its brightness tapers off sharply) is quite closely comparable in both maps (difference of less than 10\%).
As the natural-weighted image has a higher sensitivity and is therefore able to trace the circumstellar disk further out, we take the derived break radius of 87.4\,mas (12.2\,AU) as a measure of the outer radius of the circumstellar disk of the N source.  For the circumstellar disk of the S source, we take the derived break radius of 73.9\,mas (10.3\,AU) as a measure of its outer radius.  Consistent with the relative dimensions of the two disks as inferred from the Gaussian fits, the circumstellar disk of the N source is larger than that of the S source.  The ratio in outer disk sizes based on our power-law fits is 0.85, whereas that based on the FWHMs of the Gaussian fits is 0.75.


\section{Physical Parameters of Envelope}\label{Physical Parameters Envelope}

L1551\,IRS\,5 resides at the center of a flattened, rotating, and infalling envelope of dust and molecular gas \citep{Momose1998, Takakuwa2004, Chou2014}.  In C$^{18}$O \citep{Momose1998}, the centrally-condensed portion of the envelope has a size of $2380 \times 1050$\,AU as measured at the $3\sigma$ level, and a position angle for its major axis of 162\degr.  The inclination of the envelope thus determined is $\sim$64\degr.  In CS(7-6) \citep{Chou2014}, the envelope is inferred to transition to a rotationally-supported circumbinary disk (i.e., exhibiting Keplerian motion) at a radius of $\sim$0\farcs46 (projected radius of $\sim$64\,AU).  From a Keplerian fit to the spatial-kinematic structure of the circumbinary disk (assuming a geometrically-thin disk), \citet{Chou2014} infer an inclination of $\sim$60\degr\ for this disk and a position angle of $\sim$147\degr\ for its major axis.  Both the inclination and position angle for the major axis of the envelope are therefore closely comparable (within $\sim$15\degr) with the corresponding parameters for the two circumstellar disks (Table\,\ref{Fitting Parameters}).  Below, we provide a more accurate measure of the geometry of the envelope, which as we will show is even more closely aligned with the circumstellar disks than the aforementioned values indicate.    

\subsection{Inclination}\label{Inclination Envelope}
\begin{figure}
\epsscale{1.2}
\center
\plotone{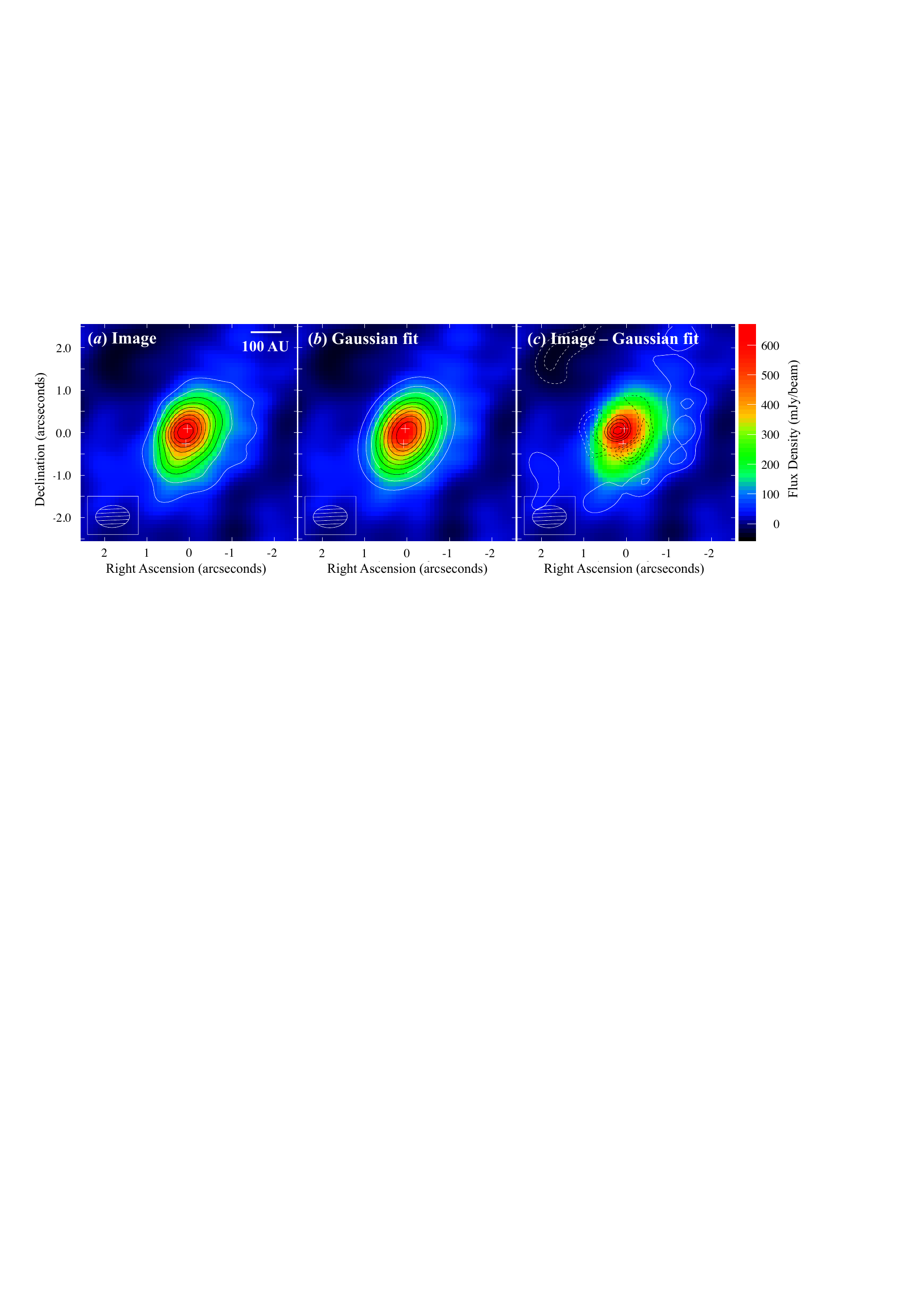}
\renewcommand{\baselinestretch}{1.3}
\vspace{-16.5cm}
\caption{Color map of the dust envelope associated with L1551\,IRS\,5 at 0.8\,mm, using uniform weighting of the data taken in the extended configuration of the SMA as obtained by \citet{Chou2014}.  The two crosses indicate the centroids of the N and S sources as measured with the VLA and reported in this manuscipt.  The synthesized beam is shown at the lower left corner of each panel.  $(a)$ Contour levels plotted at 10\%, 20\%, ... 90\% of the peak intensity of the envelope.  $(b)$ Contours levels of a two-dimensional Gaussian function fitted to the envelope, plotted at 10\%, 20\%, ... 90\% of the peak intensity.  $(c)$ Residuals after subtracting the fitted Gaussian function from the map, plotted at -50\%, -40\%, -30\%, -20\%, 30\%, 50\%, 70\%, and 90\% of the peak intensity of the strongest, positive, residual.  This residual may contain a contribution from the (more intense) circumstellar disk of the N source.  An elliptical ring-like negative residual surrounds the central positive residual, and may correspond to an inner clearing in the envelope.}
\label{Pseudodisk}
\end{figure}

In Figure\,\ref{Pseudodisk}$a$, we show a continuum map at 0.8\,mm tracing dust in the envelope around L1551\,IRS\,5.  This map was made from data obtained by \citet{Chou2014} with the SubMillimeter Array (SMA), but using only that taken in the extended configuration so as to trace the inner region of the dust envelope.  The data used was taken in 2013 January 14, close in time to our observations of L1551\,IRS\,5 with the VLA.  The positions of the two protostars (as determined from our VLA observations {and indicated by the two crosses in Figure\,\ref{Pseudodisk}}) relative to the dust envelope can therefore be accurately located without having to correct for absolute proper motion.  The emitting region traced in this map is about an order of magnitude smaller than the centrally-condensed portion of the envelope traced in C$^{18}$O by \citet{Momose1998}.  

In Figure\,\ref{Pseudodisk}$b$, we show in contours a two-dimensional Gaussian function, having the parameters listed in Table\,\ref{Parameters Pseudodisk}, fitted to the continuum source.  The residuals after subtracting this Gaussian function from the map is shown in Figure\,\ref{Pseudodisk}$c$.  
As can be seen, the fitted Gaussian function provides a good representation of the continuum source except around its central and outer eastern regions.   With a radius at half intensity along its major axis of $0\farcs72 \pm 0.01$, the continuum source extends well beyond the putative circumbinary disk (radius of $\sim$0\farcs46).  Based on its deconvolved dimensions at FWHM, we infer an inclination of $54\degr.1 \pm 1\degr.0$ for the flattened envelope, closely comparable to the inclinations of the circumstellar disks in L1551\,IRS\,5 (difference of $6\degr.0 \pm 1\degr.1$ for the circumstellar disk of the N source, and $9\degr.8 \pm 1\degr.3$ for the circumstellar disk of the S source).

Like for the circumstellar disks, when deriving the inclination of the envelope, we assumed implicitly that the envelope has a circularly symmetric radial intensity profile and is geometrically thin.  \citet{Momose1998} investigated a number of simple analytical models to reproduce the observed position-velocity (PV-) diagram measured in C$^{18}$O for the envelope around L1551\,IRS5.  They found that the observed PV-diagram is better fit by an envelope with a finite thickness rather than an infinitesimally thin disk.  For a vertical thickness to radius of $\sim$1/5 (at a given isodensity contour; their preferred model for a sheetlike envelope), the inferred inclination of the envelope is only slightly changed to $\sim$60\degr. 

\subsection{Alignment with Circumstellar Disks}\label{Alignment Envelope}
From the same two-dimensional Gaussian function fitted to the continuum source, we determined a position angle for the major axis of the flattened envelope of $156\degr.6 \pm 1\degr.3$.  The position angle of the envelope is therefore closely comparable with that of the two circumstellar disks (difference of $8\degr.3 \pm 1\degr.2$ for the circumstellar disk of the N source, and $8\degr.9 \pm 1\degr.6$ for the circumstellar disk of the S source).  Given also their comparable inclinations, 
{\it the circumstellar disks of the binary protostars in L1551\,IRS\,5 are closely parallel not just with each other but also with their surrounding flattened envelope,} providing a second constraint that we will impose on models for the formation of this system.


\subsection{Central Cavity?}
As described by \citet{Pichardo2005} and \citet{Pichardo2008}, tidal torques exerted by both protostars cause orbits in their circumbinary disk to become eccentric.  In a manner analogous to the circumstellar disks, the smallest non-intersecting orbit in the circumbinary disk defines the innermost stable orbit and hence the minimum size of a central cavity.  
In Figure\,\ref{Pseudodisk}$a$, no central depression, let alone cavity, is evident in the image of the dust emission \citep[or indeed in images of molecular-gas emission as presented by][]{Chou2014}.
At an angular resolution of $0\farcs80 \times 0\farcs52$ in Figure\,\ref{Pseudodisk}$a$, however, emission from the circumstellar disks cannot be separated from that of the circumbinary disk.  Thus, even if the circumbinary disk has a central cavity, no central depression let alone cavity would be evident in the image shown in Figure\,\ref{Pseudodisk}$a$.

The residual map of Figure\,\ref{Pseudodisk}$c$ shows a positive residual close to the centroid of the continuum source.  This positive residual is point-like, consistent perhaps with a contribution from the more intense circumstellar dust disk of the N source.  Surrounding the central positive residual is a ring-like distribution of negative residuals.  The major axis of this ring lies at the same position angle as the fitted Gaussian function.  Evidently, away from the center, the dust emission has a profile that is not as strongly centrally peaked as a Gaussian function, which otherwise provides a good fit to the bulk of the dust emission.  The presence of the ring-like negative residuals may indicate a central depression or cavity in the dust envelope.  On the other hand, the intensity profile of the envelope may simply deviate from a Gaussian function in its inner regions.

\section{Orbit}\label{Orbit}

\subsection{Relative Proper Motion}\label{Relative Proper Motion}
\begin{figure}
\epsscale{1.0}
\center
\plotone{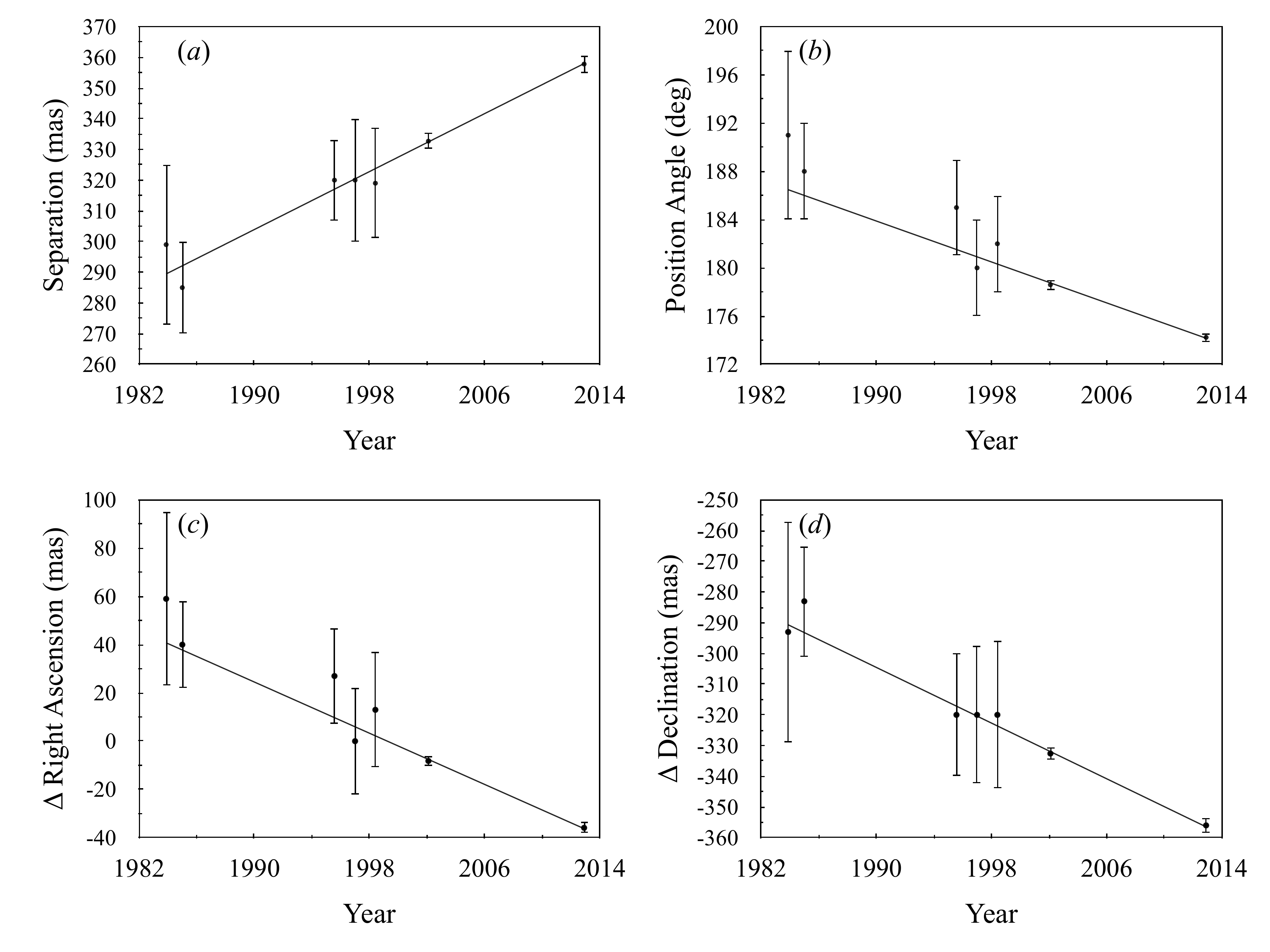}
\vspace{-1.0cm}
\renewcommand{\baselinestretch}{1.3}
\caption{Centroid location of the S source relative to the N source, showing their $(a)$ separation, $(b)$ position angle, $(c)$ relative Right Ascension, and $(d)$ relative Declination over time from 1983 to 2012.  The two most recent measurements, in 2002 \citep{Lim2006} and 2012 (this paper), were made with the full complement of the VLA at 7\,mm, and have the highest angular resolutions.  The fourth most recent measurement, in 1997 \citep{Rodriguez1998}, also was made with the VLA at 7\,mm, but only when about half of the array was equipped with the appropriate receivers.  At 7\,mm, the emission is dominated by dust in the circumstellar dust disks, but includes free-free emission from the ionized jets associated with the two sources.  The remaining measurements were made at 2\,cm, tracing free-free emission from the ionized jets \citep{Rodriguez2003a}.}
\label{proper motion}
\end{figure}

Figure\,\ref{proper motion} shows the separation and position angle of the S source relative to the N source over the last $\sim$30\,yrs.  The most recent measurement, in 2012, is that reported here, based on the centroids of the Gaussian components fitted to the circumstellar disks of the N and S sources in the robust-weighted map.  Emission from the jets is much weaker in this map than in the natural-weighted map, and is less likely to affect the determination of the centroids of the circumstellar disks.  The second last measurement, in 2002, also was made at 7\,mm with the VLA \citep{Lim2006}.  Similarly, the third measurement, in 1997, was made at 7\,mm with the VLA, although at a time when only about half of the array was equipped with the necessary receivers \citep{Rodriguez1998}.  The remaining measurements, in 1983, 1985, 1995, and 1998, were all made at 2\,cm with the VLA \citep{Rodriguez2003a}.  As can be seen, the two most recent measurements provide the strongest constraints on the relative proper motion of the binary protostars, and motivate in large part a full exploration of possible orbital solutions as described next in $\S\ref{Orbital Solutions}$.  These two measurements confirm the trend for an increasing separation between the binary protostars over time, and that the position angle of the S source as measured from the N source (anticlockwise from north) is decreasing over time (indicating clockwise orbital motion).  

The flattened envelope around L1551\,IRS\,5 has its major axis (at a position angle of $\sim$157\degr; Table\,\ref{Parameters Pseudodisk}) closely orthogonal to that of the ionized jets \citep[at position angles of about 60\degr;][]{Rodriguez2003b}  and bipolar molecular outflow \citep[at a position angle of about 45\degr;][]{MoriartySchieven1998}  from the binary prototars as projected onto the sky plane.   
If the equatorial plane of the flattened envelope is indeed orthogonal to the outflow axis (i.e., so that its eastern side is the nearer side), then the kinematics measured in molecular lines imply that the envelope is rotating in the clockwise direction.  In this case, as noted by \citet{Lim2006}, {\it the orbital motion of L1551\,IRS\,5 is in the same direction as the rotational motion of its surrounding envelope}, providing a third constraint that we will impose on models for the formation of this system.

\subsection{Orbital Solutions}\label{Orbital Solutions}
We fitted orbital solutions, {at selected orbital periods as indicated by the data points in Figure\,\ref{Best-Fit Orbits}}, to the measurements of relative proper motion shown in Figure\,\ref{proper motion} for eccentricities, $e$, of $e = 0$ (circular orbit) and $e = 0.1, 0.2, 0.4, 0.6$, and 0.8 (elliptical orbits).  Figure\,\ref{Best-Fit Orbits}$(a)$--$(b)$ shows the reduced $\chi^2$ of the best-fit orbit(s) at  {the selected} orbital {periods} for the different eccentricities considered, {the latter as indicated by the different colors with corresponding eccentricities defined at the upper right corner of Fig.\,\ref{Best-Fit Orbits}$(a)$}.  For circular orbits, there is a unique acceptable (i.e., reduced $\chi^2 \leq 1$) best-fit orbit at a given orbital period.  For elliptical orbits, we find two families of acceptable best-fit orbits at a given orbital period: (i) where the system is leaving periastron or approaching apastron (family A), plotted in the left column of Figure\,\ref{Best-Fit Orbits}; and (ii) where the system is leaving apastron or approaching periastron (family B), plotted in the right column of Figure\,\ref{Best-Fit Orbits}.  In family A (B), the system is closer to apastron (periastron) as the period decreases (increases).  As can be seen, acceptable orbital solutions exist at all the different eccentricities considered.  The orbital periods of acceptable circular orbits have a lower {bound} of $\sim$250\,yr, but no upper {bound}.  The orbital periods of acceptable elliptical orbits have both a lower and an upper bound, with a range that increases with increasing eccentricity from just $\sim$40--520\,yr 
at $e = 0.1$ to $\sim$20--5000\,yr 
at $e = 0.8$.   

Figure\,\ref{Best-Fit Orbits}$(c)$--$(d)$ shows the semi-major axis of the best-fit {orbits} at  {the selected} orbital {periods}, {with the different colors again corresponding to orbits having the different eccentricies considered}.  Over a broad range of orbital periods ($\sim$20--5000\,yr), the orbital semimajor axis spans the relatively narrow range $\sim$0\farcs3--1\farcs5 ($\sim$40--210\,AU).  In family A, the semi-major axis is nearly independent of eccentricity at a given orbital period.  In family B, the semi-major axis increases with eccentricity at a given orbital period.  Figure\,\ref{Best-Fit Orbits}$(e)$--$(f$) shows the corresponding total mass of the binary system as a function of orbital period, {with the different colors corresponding to the different orbital eccentricies considered}.  For circular obits, the total binary mass attains a minimum value of $\sim$0.5\,$\rm M_\sun$ at orbital periods in the range $\sim$600--1000\,yr.  For elliptical orbits in family A, the total binary mass is confined to a relatively narrow range spanning $\sim$0.5--1.0\,$\rm M_\sun$ over a broad range of orbital periods spanning $\sim$30--2000\,yr.  For elliptical orbits in family B, the total binary mass increases with eccentricity, spanning the range from $\sim$1\,$\rm M_\sun$ at $e=0.1$ to $\sim$9\,$\rm M_\sun$ at $e=0.8$.

Figure\,\ref{Best-Fit Orbits}$(g)$--$(h)$ shows the inclination of the orbital plane for the best-fit {orbits} at the selected orbital {periods} and {eccentricities}.  We note that the inclination of the orbital plane is degenerate insofar that, without knowledge of whether the N or S source is approaching us, we do know which of the nodes correspond to the ascending and which to the descending node. (In the same way, we do not know which sides of the circumstellar disk, along its minor axis, are the near and far sides.)  In the same panels, the inclinations of the circumstellar disks derived from the natural-weighted maps ($\S\ref{Inclination Circumstellar Disks}$) are indicated by shaded strips (the widths of the strips reflecting the $\pm 1\sigma$ uncertainty in the derived inclinations; Table\,\ref{Fitting Parameters}). Locations where the shaded strips intersect solutions for the orbital inclination indicate orbits that have the same inclination as one of the circumstellar disks.  For circular orbits, the orbital inclination is systematically higher than the disk inclinations by  $\sim$15\degr--25\degr\ {(based on their formal values)} over the range in orbital periods spanning $\sim$300--1000\,yr.  For elliptical orbits in family A, the orbital inclination differs from the disk inclinations by no more than about $\pm 20\degr$ {(based on their formal values)}, a difference that is usually much smaller than $3\sigma$.  For elliptical orbits in family B, the orbital inclination is systematically higher than the disks inclinations, increasing with eccentricity from $\sim$20\degr\ at $e = 0.1$ to $\sim$35\degr\ at $e = 0.8$.

Figure\,\ref{Best-Fit Orbits}$(i)$--$(j)$ shows the position angle of the line of nodes  {for the best-fit orbits at the selected orbital periods and eccentricities}. The shaded strips indicate the position angles of the major axes of the circumstellar disks for the N and S sources, with widths reflecting the $\pm 1\sigma$ uncertainties, as inferred from the natural-weighted maps (Table\,\ref{Fitting Parameters}).  Locations where the shaded strips intersect solutions for the position angle of the line of nodes indicate orbits that cut through the plane of the sky at the same position angle as the circumstellar disks.  For circular orbits, the position angle of the line of nodes  {differ from the position angle of the disk major axes by no more than $\sim$20\degr\ (based on their formal values)}.  
For elliptical orbits in family A, the position angle of the line of nodes differ by no more than $\sim$30\degr\ from the position angle of the disk major axes over the eccentricity range $0 \leq e \leq 0.4$.  
This difference can be as large as $\sim$50\degr\ at $e = 0.6$, and as large as $\sim$70\degr\ at $e = 0.8$.  As we will show in $\S\ref{Constraints Circumstellar Disks}$, however, constraints on the circumstellar disks sizes as imposed by tidal truncation limit acceptable orbits with $e = 0.6$ in family A to those where the position angle of the line of nodes axis are quite closely aligned (within $\sim$20\degr) with the disk major axes.  
For elliptical orbits in family B, the position angles of the line of nodes differ by no more than $\sim$20\degr\ from the position angle of the disk major axes.  

In summary, although acceptable orbital solutions can be found over a broad range of eccentricities, all (omitting those in family A having eccentricities $e \geq 0.6$ for the reason mentioned above) imply a relatively close alignment (mostly $\lesssim 25\degr$) between the circumstellar disks and the orbital plane.  {\it The circumstellar disks of the binary protostars in L1551\,IRS\,5 are therefore quite closely aligned with the orbital plane, possibly essentially coplanar although more probably tilted by up to typically $\sim$25\degr\ from the orbital plane}.  This condition provides the fourth constraint that we will impose on models for the formation of L1551\,IRS\,5.  

\begin{figure}
\epsscale{1.0}
\center
\vspace{-0.8cm}
\plotone{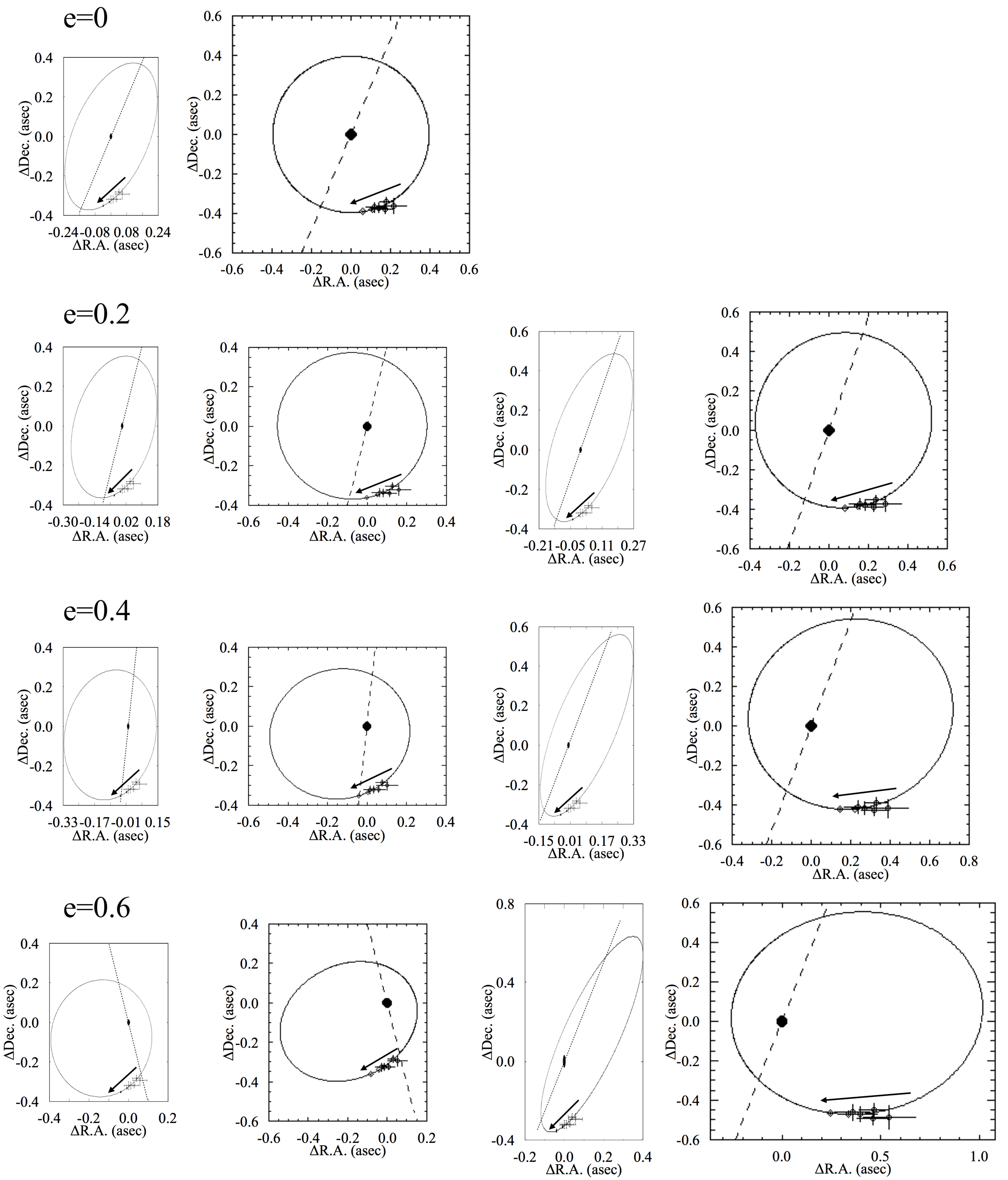}
\vspace{-0.4cm}
\renewcommand{\baselinestretch}{1.0}
\caption{\small Examples of best-fit orbits all having a common orbital period of 500\,yr.  First two columns are for Family A, and last two columns are for Family B.  The first and third columns are projected orbits, and the second and fourth columns deprojected orbits.  Dotted lines correspond to the line of nodes.  The same lines are drawn as dashed lines in the deprojected orbits.  The points correspond to the position of the S source relative to the N source from measurements of the relative proper motion, and have arm lengths corresponding to the $\pm 1\sigma$ uncertainty.  The arrows indicate the direction of motion of the S source.}
\label{Examples Best-Fit Orbits}
\end{figure}

Figure\,\ref{Examples Best-Fit Orbits} shows examples of best-fit orbits all having a common orbital period of 500\,yr.  {The first and third columns show the projected orbits for family A and B, respectively.  The second and fourth columns show the deprojected orbits (as seen along the orbital axis) for family A and B, respectively.}  At this orbital period, the orbit for $e=0.8$ (not shown) is very similar to that for $e=0.6$ in a given family.  
As can be seen, for all the orbits shown, the available measurements of relative proper motion span only a small fraction (less than one-tenth) of a complete orbit, explaining why the orbital eccentricity is poorly constrained.  Despite this obvious limitation, the available measurements already constrain the best-fit orbits to being quite closely if not essentially coplanar with the circumstellar disks.

\subsection{Orbital Constraints}\label{Orbital Constraints}

\subsubsection{Constraints from Circumstellar Disks}\label{Constraints Circumstellar Disks}


In a coplanar binary system, the radii of the two circumstellar disks combined obviously cannot be larger (and, indeed, must be significantly smaller) than the binary separation.  Furthermore, for elliptical orbits, \citet{Pichardo2005} find that the radius of the outermost non-intersecting orbit in a circumstellar disk varies little with orbital phase, generally by only about 10\%.  Thus, in a coplanar binary system, the orbital separation at periastron (closest approach) sets a limit on the maximal possible sizes (as imposed by tidal truncation) for the circumstellar disks of the binary components.  The tidally-truncated radius of each circumstellar disk also depends on the mass ratio of the binary companions, with the more massive companion having a larger tidally-truncated radius for its circumstellar disk.  For a given mass ratio, the inferred sizes of the circumstellar disks in L1551\,IRS\,5 can therefore be used to rule out orbits with too small a periastron separation, a point we previously touched upon in \citet{Lim2006} and fully explore below.

To begin, we assume equal protostellar masses (i.e., $q=1$) and hence equal maximum possible disk sizes.  Note that \citet{Pichardo2005} define $q = m_s/(m_p+m_s)$, where $m_p$ is the mass of the primary and $m_s$ the mass of the secondary, whereas we define $q = m_s/m_p$ as is commonly used.  The {data} points in Figure\,\ref{Orbital Constraints}($a$)--($d$) indicate the predicted tidally-truncated radii (averaged over an orbit) of the circumstellar disks (identical for the two protostars) for the best-fit orbits having the different eccentricities considered, {the latter as indicated in the different colors}.  Like in Figure\,\ref{Best-Fit Orbits}, the left column corresponds to the best-fit orbits in family A and the right column those in family B.  At a given orbital period, the predicted tidally-truncated radius  decreases as the orbital eccentricity increases, as is expected.  The break radius for the circumstellar disk of the N source is indicated by the horizontal line in Figure\,\ref{Orbital Constraints}($a$--$b$), and that for the circumstellar disk of the S source by the horizontal line in Figure\,\ref{Orbital Constraints}($c$--$d$).  {The data points lying above the horizontal line in each panel therefore correspond to orbits where the predicted tidally-truncated radius of the circumstellar disk is larger than its observed break radius, whereas the data points lying below the horizontal line to orbits where the predicted tidally-truncated radius of the circumstellar disk is smaller than its observed break radius.}  The curves {joining the data points} indicate the best-fit orbits for which  {the predicted tidally-truncated radius is, to within an uncertainty of $3\sigma$, no smaller than the observed break radius.}  As can be seen,  {the} condition {that the predicted tidally-truncated radius cannot be smaller than the observed break radius} rules out an increasingly larger range in orbital periods as the orbital eccentricity increases.  For $q=1$, the larger circumstellar disk of the N source imposes a more stringent constraint than the smaller circumstellar disk of the S source.  For the former disk, the aforementioned condition can be met over nearly the entire range of orbital periods spanned by the best-fit orbits in family A only for $e \leq 0.1$.  This condition can be met over the entire range of orbital periods spanned by the best-fit orbits in family B only for $e \leq 0.2$.  For both families, this condition is satisfied over only a limited range in orbital periods spanned by the best-fit obits for $0.4 \leq e \leq 0.6$, and cannot be at all satisfied for $e = 0.8$. 

Whether fitted by a Gaussian or a NUKER profile, the circumstellar disks of the N and S sources have different widths as measured at their FWHM (Table\,\ref{Fitting Parameters}) or their break radii (Table\,\ref{Parameters NUKER fits}).  This difference raises the possibility that the two protostars in L1551\,IRS\,5 have different masses, resulting in different tidally-truncated radii for their circumstellar disks.  Based on the ratio in their break radii of about 0.85, the mass ratio of the two protostars according to the calculations of \citet{Pichardo2005} is about 0.72.  For $q = 0.72$, the {data} points in Figure\,\ref{Orbital Constraints}($e$)--($f$) indicate the predicted tidally-truncated radius for the circumstellar disk of the N source, and those in Figure\,\ref{Orbital Constraints}($g$)--($h$) the predicted tidally-truncated radius for the circumstellar disk of the S source, for the best-fit orbits having the different eccentricities considered.  Like before, the left column corresponds to the best-fit orbits in family A and the right column those in family B.  As can be seen, at a given orbital period, the tidally-truncated radius of the circumstellar disk for the more massive northern protostar is larger than that for the less massive southern protostar.  Compared with $q=1$, the tidally-truncated radius for the circumstellar disk of the N source is larger (except for $e=0.6$, due to the complex manner in which the tidally-truncated radius changes with orbital phase), whereas that for the circumstellar disk of the S source is smaller.  For both circumstellar disks, the changes can be quite small at high eccentricities so as to be unnoticeable.  The break radius for the circumstellar disk of the N source is indicated by a horizontal line in Figure\,\ref{Orbital Constraints}($e$--$f$), and that for the circumstellar disk of the S source by a horizontal line in Figure\,\ref{Orbital Constraints}($g$--$h$).  As before, the curves {joining the data points} indicate the best-fit orbits for which  {the predicted tidally-truncated radius is, to within an uncertainty of $3\sigma$, no smaller than the observed break radius.}  Coincidentally, this condition rules out nearly the same range in orbital periods at a given eccentricity for both protostars as the circumstellar disk of the N source in the case where $q=1$.  

In making the aforementioned comparisons, we have assumed that the break radii of the circumstellar disks  {provide a good measure of the lower bound on} their tidally-truncated radii.  We now consider situations where the break radii of the circumstellar disks may be driven to either extend beyond or lie within their theoretically predicted  tidally-truncated radii.   First, to continue to grow, both protostars must accrete through their circumstellar disks, which in turn must be replenished by material from the surrounding envelope.  Accretion of infalling matter onto the circumstellar disks allows the disks to extend, albeit temporarily (within a disk dynamical timescale) if the accretion is intermittent, beyond their theoretically predicted  tidally-truncated radii.  Theoretical simulations that assume circular coplanar orbits find that accretion onto the circumstellar disks of binary protostars is highly unequal, with some models favouring accretion onto the circumstellar disk of the secondary component \citep{Artymowicz1983, Bate1997, BateBonnell1997} whereas others favouring the opposite \citep{Ochi2005}. For eccentric coplanar systems, \citet{Artymowicz1996} find that accretion is stronger onto the circumstellar disk of the secondary component.  A recent study by \citet{Dunhill2015}, however, paints a more complicated picture for accretion in eccentric coplanar systems.  They performed hybrid SPH/N-body simulations of a binary system having a moderately high eccentricity (e = 0.6) and a moderately small mass ratio ($q = 0.64$) accreting from its circumbinary disk.  The binary system clears a large cavity in its circumbinary disk, and drives elliptical orbits at the cavity edge.  Inflow from the circumbinary disk onto a given circumstellar disk is strongly modulated with orbital phase, such that accretion is strong only during a relatively short interval centred just before ($\sim$0.1 orbital phase from) periastron.   The binary system also drives precession in the elliptical orbits at the cavity edge.  Accretion onto the circumprimary disk is much stronger than that onto the circumsecondary disk for half of this precession period (which is more than an order of magnitude longer than the binary orbital period), and vice versa.  Averaged over time, accretion onto the circumprimary disk is similar to or slightly higher than that onto the circumsecondary disk, opposite to that found by \citet{Artymowicz1996}.  Despite their different predictions, all these studies find that accretion, when ongoing, is stronger onto one than the other circumstellar disk.  Thus, at least one of the circumstellar disks in L1551\,IRS\,5 is likely to have a size that is not affected by recent accretion. (We note that rotational periods at the predicted tidally-truncated radii in the circumstellar disks of L1551 IRS5 are, for $q \sim 1$, considerably shorter than the orbital periods of the acceptable best-fit orbits.  {In other words, dynamical times in both circumstellar disks are considerably shorter than the binary orbital period.}) 

Elliptical orbits in the circumstellar disks may be damped by gas pressure,  {providing an alternative means by which circumstellar disks can extend beyond}  {their} largest non-intersecting orbit.  If the dust-to-gas ratio is constant and the gas well coupled to the dust at all radii in the circumstellar disks, then the gas pressure $\rho_{gas} \propto r^{-{\Gamma}}$.  Equating $\Gamma = \gamma$, then given the inferred power-law index of $\gamma \sim 0.60$ for the circumstellar disk of the N source and $\gamma \sim 0.81$ for the circumstellar disk of the S source (Table\,\ref{Parameters NUKER fits}), the gas pressure only decreases modestly with increasing radius.  Although on this basis we cannot rule out that gas pressure {may} damp elliptical orbits in the outer regions of the circumstellar disks, we note that the break radii of both circumstellar disks are closely comparable with their predicted tidally-truncated radii for the best-fit circular (for orbital periods $< 1000$\,yr) or low eccentricity ($e \leq 0.2$) orbits.  It seems to us most unlikely  that, {if L1551\,IRS\,5 has has a circular or low-eccentricity orbit}, gas pressure conspires to extend both {its} circumstellar disks  so as to closely match {their} predicted tidally-truncated radii.  

Rather than extending beyond their predicted tidally-truncated radii, it is more likely that one or both circumstellar disks do not extend to their tidally-truncated radii because of depletion by accretion onto their respective protostars.  In addition, or alternatively, magnetic braking can effectively shrink the circumstellar disks through the efficient outwards transfer of angular momentum \citep{Machida2011}.  The possibility that the inferred break radii of the circumstellar disks are significantly smaller than their predicted tidally-truncated radii rules out even more of the best-fit orbits derived for L1551\,IRS\,5 ({i.e., the actual tidally-truncated radii of the circumstellar disks lie above the horizontal lines drawn in} Figure\,\ref{Orbital Constraints}), thus rejecting most if not all orbits having relatively high eccentricities ($e \geq 0.4$).  In short, the break radii inferred for the circumstellar disks of the binary protostars in L1551\,IRS\,5 strongly favor orbits having relatively low eccenticities ($e \lesssim 0.2$).

\subsubsection{Constraints from Envelope}\label{Constraints Envelope}

A central depression is seen in the circumbinary dust disk of the binary protostellar system L1551\,NE \citep{Takakuwa2013, Takakuwa2014}, attesting to the possibility of a central cavity cleared by tidal forces from the protostars.  To explore what constraints a central cavity in the circumbinary disk or envelope can place on the orbit of L1551\,IRS5, let us take the FWHM for the flattened envelope of $0\farcs72 \pm 0\farcs01$ as a firm upper limit for the size of a central cavity.  The {data} points in Figure\,\ref{Orbital Constraints}($i$)--($j$) indicate the predicted minimal radius of a central cavity for the best-fit orbits having the different eccentricities considered.  As a reminder, the left column of this figure is for orbital solutions in family A, and the right column those in family B.  The upper limit on the radius of a central cavity in the envelope of L1551\,IRS\,5 is indicated by the horizontal line.  The curves {joining the data points} indicate the best-fit orbits for which  {the predicted minimum size of a central cavity in the envelope is, to within an uncertainty of $3\sigma$, no larger than the FWHM of the dust envelope.}  Among the best-fit circular orbits, only those with orbital periods $\lesssim 720$\,yr are not ruled out at the $\geq 3\sigma$ level.  Among the best-fit orbits having $e=0.1$, only a relatively narrow range of orbital periods spanning $\sim$470--510\,yr in family A and $\sim$390--410\,yr in family B are not ruled out.  For those having $e=0.2$, all in family A are ruled out and only a very narrow range of orbital periods spanning $\sim$346--366\,yr in family B are not ruled out.  All the best-fit orbits having $e=0.4$ are ruled out.  For those having $e=0.6$, only a relatively narrow range of orbital periods spanning $\sim$300--530\,yr in family A are not ruled out, whereas all in family B are ruled out.  All best-fit orbits having $e=0.8$ are ruled out.  Thus, just like the constraints imposed by the break radii of the circumstellar disks, the upper limit paced on the size of a central cavity in the envelope strongly favors orbits having relatively low eccentricities ($e \lesssim 0.2$)

Envelopes around binary protostellar systems or their circumbinary disks, however, need not necessarily develop a central cavity as large as that imposed by tidal truncation.  For example, inflow from the surrounding envelope or circumbinary disk can extend the envelope or circumbinary disk inwards to a radius smaller than that imposed by tidal truncation.  In addition, just like in the circumstellar disks, elliptical orbits in the circumbinary disk can be damped by gas pressure, thus changing (decreasing) the radius of the smallest non-intersecting orbit.  In principle, one or both effects can extend the circumbinary disk inwards as far as its centrifugal radius (if any), where gravitational and centrifugal forces are balanced.  Even if there is a central clearing in the circumbinary disk, material is predicted to penetrate into this clearing through the outer Lagrangian points of the binary system to accrete onto the circumstellar disks.  As a consequence, any central clearing in the circumbinary disk is, at least sporadically, not entirely devoid of matter.

As mentioned in $\S\ref{Physical Parameters Envelope}$, the envelope is inferred to transition to a rotationally-supported circumbinary disk (i.e., exhibiting Keplerian motion) at a radius of $\sim$0\farcs46 (projected radius of $\sim$64\,AU).  From a Keplerian fit to the spatial-kinematic structure of the circumbinary disk, \citet{Chou2014} infer a binary mass of $\sim$$0.5 \rm \, M_{\sun}$.  Such a binary mass rules out orbits having eccentricities $e > 0.2$ in family B, and constrains the orbital period to the range $\sim$200-2000\,yrs or narrower in family A.

\section{Discussion}\label{Discussion}

\subsection{Observational Constraints on Fragmentation Models}\label{Constraints on Fragmentation Models}
We begin by summarizing the constraints imposed on models for the formation of L1551\,IRS\,5 as derived in $\S\ref{Physical Parameters Circumstellar Disks}$--$\S\ref{Orbit}$. 
From the measured geometry of its circumstellar disks and its surrounding envelope:

\begin{itemize}

\item the circumstellar disks of the binary protostars are very closely parallel

\item the circumstellar disks also are very closely parallel to their surrounding flattened envelope  

\end{itemize}

\noindent Based on the measured relative proper motion of the binary protostars and published measurements of the kinematics of its surrounding envelope:

\begin{itemize}

\item the orbital motion is in the same direction as the rotational motion of the surrounding envelope


\end{itemize}

\noindent {The close geometrical and dynamical relationship between the circumstellar disks and surrounding flattened envelope suggests that the angular momentum of material that gave rise to the protostellar system and that comprising its parental core are aligned.}

\noindent Finally, although orbits spanning a wide range of eccentricities provide acceptable fits to the available measurements of the relative proper motion for the binary protostars in L1551\,IRS5, theoretical predictions for limits in the sizes of the circumstellar disks as imposed by tidal truncation rule out the vast majority of best-fit orbits having moderate eccentricites ($0.4 \lesssim e \lesssim 0.6$) and all best fit orbits having high eccentricities ($e \gtrsim 0.8$).  For the remaining best-fit orbits:

\begin{itemize}

\item in the majority of cases, there is a systematic tilt (by up to typically $\sim$25\degr) between the circumstellar disks and the binary orbital plane; nevertheless, there is a clear preference for a relatively close alignment if not coplanarity


\end{itemize}


\subsection{Predictions of Fragmentation Models}\label{Fragmentation Models}
As explained in $\S\ref{Introduction}$, current models invoke either {local (small-scale)} turbulence in or the bulk rotation of cores to drive fragmentation.  In cores that have little or no bulk rotation, turbulence 
introduces velocity and density inhomogeneities that can seed and drive the growth of multiple density perturbations to become self gravitating \citep[e.g.,][]{Bate2002, Bate2003, Bate2005, DelgadoDonate2004a, DelgadoDonate2004b, Goodwin2004a, Goodwin2004b, Goodwin2006}.  Thereafter, dynamical interactions between the fragments can play an important role in determining the final properties of the binary system (e.g., number and masses of component stars, together with their orbital separation and eccentricity).  
Multiple fragments produced in different turbulent cells are predicted to exhibit random orientations between the circumstellar disks or spin axes of the binary components, and no particular relationship between the circumstellar disks and orbital plane or the spin and orbital axes.  If multiple fragments are produced in a common region where turbulence conspires to create local angular momentum, however, the binary system thus assembled can exhibit quite well aligned circumstellar disks and orbital plane or spin and orbital axes.

Rather than through turbulence, the large-scale ordered rotation of the core also can drive dynamical instabilities to induce fragmentation during collapse. 
In such models, conservation of angular momentum forces cores to become increasingly flattened as they collapse.  As a result, a disequilibrium disk-like (i.e., flattened and rotating) structure forms at the center of the core.  The central region of the core can become especially flattened if magnetic fields are invoked to direct infalling matter onto the mid-plane of the disk-like structure; the resulting structures closely resemble, at least morphologically, rotationally-supported disks, and are therefore referred to as pseudodisks \citep{Galli1993a, Galli1993b}.  By introducing an initial density or velocity perturbation, the large-scale ordered rotation of the core can drive dynamical instabilities in the form of a spiral, bar, or ring in its central flattened region \citep{Matsumoto2003, Cha2003, Machida2008}.  The pattern of the dynamical instability does not depend on the nature of the initial velocity or density perturbations.  Furthermore, because the perturbations introduced can either promote or hinder fragmentation, in the latter case through the effective removal of angular momentum by gravitational torques from the resulting dynamical instability, increasing the amplitude of the initial perturbation does not necessarily increase the likelihood of fragmentation.  Fragments form in localised regions of the resulting dynamical instabilities that are gravitationally unstable (according to the Toomre criterion) and have masses exceeding the local Jeans mass.  Theoretical considerations suggest that, when driven by the large-scale ordered rotation of the core, fragmentation most likely occurs during the adiabatic phase (when a first hydrostatic core forms and grows) or the protostellar phase \citep[see arguments in][]{Machida2008}.

Binaries that form through the rotation fragmentation of disk-like structures should naturally exhibit a close alignment between the spin axes of the component stars.  Because fragments can be produced at different heights from and perhaps even on opposite sides of the mid-plane, however, the circumstellar disks and orbital planes of the resulting binary system need not be closely aligned but can span a  range of angles.  
The mass ratio and orbital parameters (i.e., orbital separation and eccentricity) of such binary systems depend not only on the initial mass ratios and orbits of their parental fragments, but also how these fragments interact with each other (especially in systems initially comprising three or more fragments) as well as with the surrounding material from which they accrete.  

\subsection{Formation of L1551\,IRS\,5}

As explained above, both turbulent- and rotationally-driven fragmentation can produce binary systems in which the circumstellar disks are parallel with each other, as well as quite closely parallel if not coplanar with the binary orbit, satisfying some of the observed physical properties of L1551\,IRS\,5.  The crucial distinction between the two fragmentation models, however, is the physical relationship between the binary system and its surrounding bulk envelope, constituting the remnant parental core.  In models that invoke turbulent fragmentation, even in the situation where multiple fragments are produced in a common region where turbulence conspires to create local angular momentum (and thus produce an aligned binary system), the binary system need not orbit in the same direction as the rotation of the bulk envelope, let alone have circumstellar disks aligned with any flattening of the bulk envelope.  On the other hand, in models that invoke rotational fragmentation, the binary system will of course orbit in the same direction as the rotation of its surrounding envelope, and its circumstellar disks aligned with the surrounding flattened envelope.  The observed physical properties of L1551\,IRS\,5 therefore directly point to its formation through rotational fragmentation.  In support of this argument, both the measured rotational velocity and the high degree of flattening exhibited by the envelope around L1551\,IRS\,5 suggest that its parental core possessed considerable angular momentum.

\begin{figure}
\epsscale{1.2}
\center
\plotone{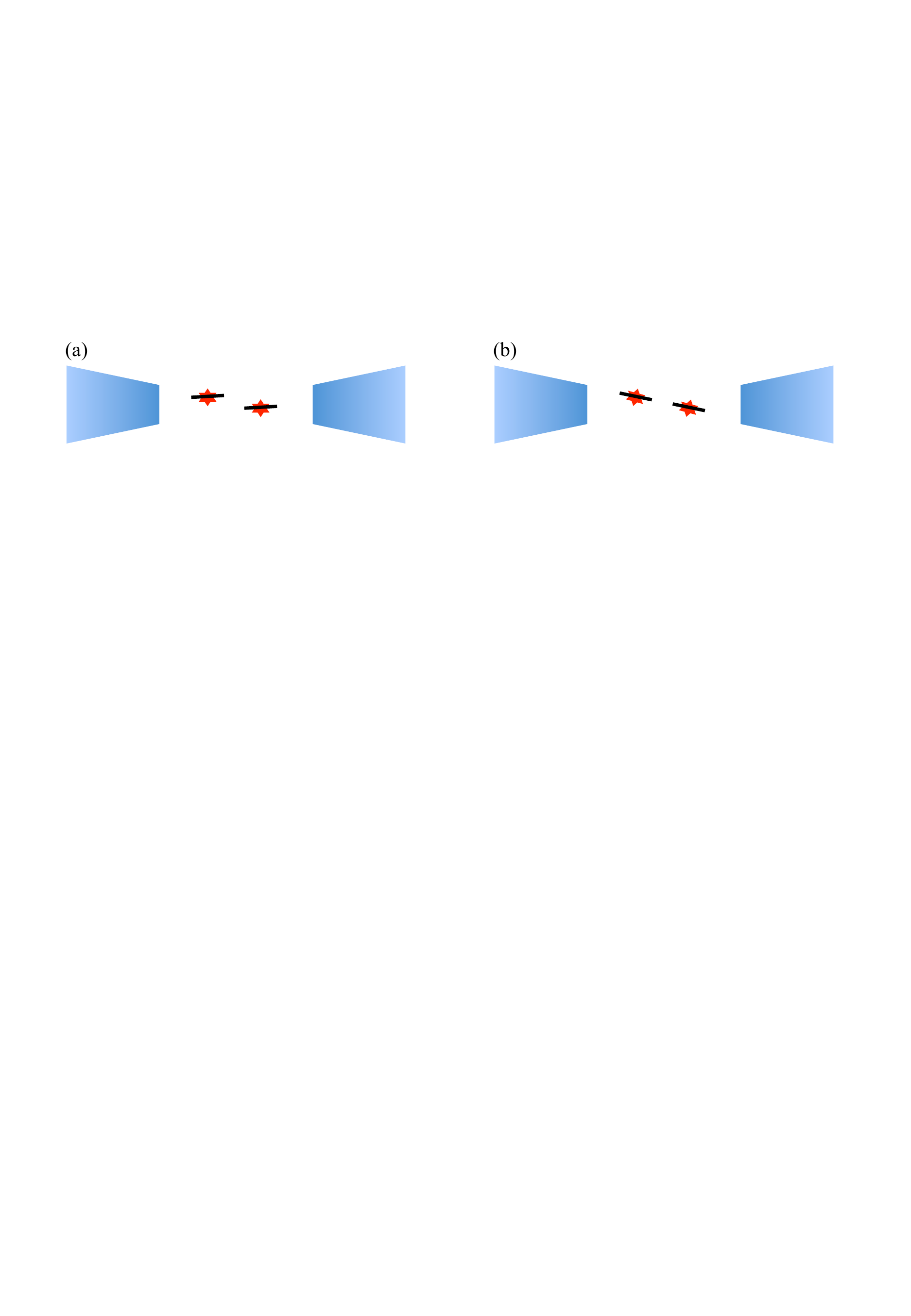}
\renewcommand{\baselinestretch}{1.3}
\vspace{-19.5cm}
\caption{Schematic illustrating a binary protostellar system formed through rotationally-driven fragmentation of its flattened parental core.  {The red symbols indicated the binary protostars, the thick black lines their circumstellar disks, and the blue-shaded regions a vertical cross section through their surrounding flattened envelope.}  (a) Fragments that gave rise to the binary protostars formed at different heights from the midplane of the core, resulting in a binary system having circumstellar disks closely parallel with each other as well as with the surrounding envelope, but tilted from the orbital plane.  (b) Subsequent tidal interactions between one protostar and the circumstellar disk of the other align both circumstellar disks with the orbital plane.  Now, however, the mutual plane of the system is tilted from the surrounding envelope.  Situation (a) most probably applies to L1551\,IRS\,5.}
\label{Fragmentation}
\end{figure}

As mentioned above, even when formed by rotational fragmentation, the circumstellar disks of the resulting binary protostars, although parallel with each other, need not be {accurately} aligned with the orbital plane.  \citet{Bateetal2000} argue that tidal interactions can quite rapidly align the circumstellar disks of binary protostars with their orbital planes, such that low-mass binary protostellar systems with orbital separations $\lesssim 100$\,AU should have coplanar disks and orbital planes.  If this process had indeed driven the circumstellar disks of L1551\,IRS\,5 towards close alignment with the binary orbital plane, the same process would have led to a misalignment between the circumstellar disks/orbital plane and the surrounding flattened envelope.  Consider the situation where the two protostars formed at different heights from, and perhaps even on opposite sides of, the midplane of a flattened core, as shown in the schematic diagram of Figure\,\ref{Fragmentation}($a$).  At birth, the circumstellar disks of the two protostars are therefore closely parallel with their surrounding flattened envelope, but not coplanar with each other or the orbital plane.  If the circumstellar disks are then brought into alignment with the orbital plane by tidal interactions, the mutual plane of the binary system now becomes misaligned with the surrounding flattened envelope, as shown in Figure\,\ref{Fragmentation}($b$).  By contrast, we find that the circumstellar disks of L1551\,IRS\,5 are very closely parallel to their surrounding flattened envelope, but probably misaligned by up to typically $\sim$25\degr\ from the orbital plane, a geometry more akin to that shown in Figure\,\ref{Fragmentation}($a$).  The probable misalignment between the circumstellar disks/flattened envelope and the binary orbital plane may therefore reflect the formation of the binary fragments in different planes sharing the same rotational axis.

By imposing the condition that the inferred break radii of the circumstellar disks cannot be larger than their theoretically predicted tidally-truncated radii, only a very narrow range of orbital periods, if any, are permitted for the best-fit orbits having eccentricities $e \gtrsim 0.4$, as explained in $\S\ref{Constraints Circumstellar Disks}$.  Thus, most probably, L1551\,IRS\,5 has a circular or low-eccentricity orbit.  Early in the protostellar and perhaps even first core phase when fragmentation occurs, the fragments have a very low mass compared with their surrounding envelope (parental core).  In the situation where, at apastron, the orbital velocity of the fragments is smaller than the rotational velocity of their surrounding envelope (in the immediate vicinity), tidal interactions between the envelope and the fragments transfer both angular momentum and energy from the envelope to the binary.  This action results in an increase in the orbital velocity and hence separation during the subsequent passage through periastron, and consequently a reduction in the orbital eccentricity.  Similarly, at periastron, if the orbital velocity of the fragments is larger than the rotational velocity of their surrounding envelope (in the immediate vicinity), tidal interactions between the fragments and the envelope transfer both angular momentum and energy from the binary to the envelope.  This action results in a decrease in the orbital velocity and hence separation during the subsequent passage through apastron, and once again a reduction in the orbital eccentricity.    Thus, the probable low orbital eccentricity of L1551\,IRS\,5 may have been imprinted at birth.

\subsection{Future Evolution}
The total mass of L1551\,IRS\,5 is at least $\sim$$0.5 {\rm \ M_\sun}$ (based on the best-fit orbits), compared with a mass for its envelope of about $0.1 {\rm \ M_\sun}$ \citep{Momose1998}.  Thus, during its remaining protostellar phase, L1551\,IRS\,5 is not likely to accrete more than $\sim$20\% of its present total mass and perhaps much less.  Thus, both the total mass and the mass ratio of the binary system should be largely preserved as it evolves onto the pre-main-sequence and then the main-sequence.  Whether or not the orbital parameters of L1551\,IRS\,5 will evolve significantly during its remaining protostellar phase is not so clear.  In the situation where the binary system dominates in mass over its circumbinary disk (so that, even at apastron, the binary orbits faster than the circumbinary disk rotates), as must be the case here, theoretical simulations show that tidal interactions between the binary system and its (coplanar) circumbinary disk result in a transfer of both angular momentum and energy from the binary to its circumbinary disk \citep{Artymowicz1991, Bate2000}.  The effect is to drive up the orbital eccentricity, opposite to the situation described above where the envelope dominates in mass over the binary fragments.  By preserving the apastron distance, $a(1+e)$, the effect also is to decrease the orbital semimajor axis.  

Thus, L1551\,IRS\,5 will likely evolve into a solar-mass binary system that has roughly equal companion masses, quite closely coplanar circumstellar disks and orbital planes (especially if tidal interactions between one protostar and the circumstellar disk of the other protostar bring their circumstellar disks further into alignment with the orbital plane), and hence presumably also closely aligned stellar spin and orbital axes.  Unless its orbital parameters are dramatically altered during its subsequent evolution, L1551\,IRS\,5 will retain a likely circular or low eccentricity orbit with a semimajor axis of many tens to perhaps a few hundred AU, and a corresponding orbital period of many hundreds to perhaps a few thousand years.  The physical properties that we project L1551\,IRS\,5 will have in its final state closely resemble those observed for binary systems on the pre-main-sequence and main-sequence having similar masses and orbital separations \citep[e.g., see review by][]{Duchene2013}.  Such systems do not show a preferred mass ratio or orbital eccentricity, but do exhibit preferentially aligned circumstellar disks on the pre-main-sequence and preferentially aligned stellar spin and orbital axes on the main-sequence.
For example, the circumstellar disks of T\,Tauri star binaries having projected orbital separations spanning the range $\sim$100--1000\,AU are preferentially aligned to within a few tens of degrees \citep[e.g.,][]{Wolf2001, Jensen2004, Monin2006}.  
Main-sequence solar-type {binaries} having orbital separations smaller than about 30--40\,AU (close to the median orbital separation) have approximately aligned spin and orbital axes, whereas those with larger orbital separations commonly show misaligned spin and orbital axes \citep{Hale1994}.  Many of the systems that are closely aligned may have formed in a manner similar to L1551\,IRS\,5, rather than having being brought into alignment by internal or external interactions.

\section{Summary}\label{Summary}
We have mapped the circumstellar dust disks of the binary protostars L1551\,IRS\,5 at an unprecedented sensitivity and at a high angular resolution at 7\,mm with the VLA.  This map also provides a precise second measurement of the relative proper motion of the two protostars separated by a decade; combined with earlier observations, measurements of their relative proper motion now span an interval of $\sim$30\,yr.  By fitting a two-component, two-dimensional Gaussian function to each source, with one component corresponding to dust emission from the circumstellar disk and the other to free-free emission from an ionized jet projected against the disk, we found that:


\begin{itemize}

\item the circumstellar disk of the northern (N) source has an inclination of $48\degr.1 \pm 0\degr.5$ and that of the southern (S) source an inclination of $44\degr.3 \pm 0\degr.8$, so that both disks have nearly the same inclinations

\item the circumstellar disk of the N source has a position angle for its major axis of $148\degr.2 \pm 0\degr.7$ and that of the southern (S) source a position angle for its major axis of $147\degr.8 \pm 3\degr.6$, and so both disks have essentially identical position angles for their major axes

\end{itemize}

\noindent Thus, {\it both circumstellar disks are closely parallel.}

To estimate the sizes of the two circumstellar disks, we fitted a double power-law profile, known as the NUKER function, to the jet-subtracted image (i.e., circumstellar disk) of each source.  In this way, we mimic what would be expected if the disks have power-law surface density and temperature profiles (represented by the inner power-law), and a relatively sharp cutoff at their outer edges imposed by tidal interactions with their companion protostars (represented by the outer power-law).  We found that, within the uncertainties, fitting NUKER functions yielded the same inclinations and positions angles for the major axes of the two disks as were found by fitting Gaussian functions.  Fixing these parameters at their formal values as mentioned above, from the NUKER fits we found that:

\begin{itemize}

\item the circumstellar disk of the N source tapers off sharply at a radius of $\sim$12.1\,AU and that of the S source at a radius of $\sim$10.3\,AU, values we take as measures of the tidally-truncated radii of the two circumstellar disks

\end{itemize}

We found that both circular and elliptical orbits, the latter having eccentricities as high as $e = 0.8$ (the highest that we tried), provide acceptable fits to available measurements for the relative proper motion of the binary protostars.  Nevertheless, for all the best-fit acceptable orbits:

\begin{itemize}

\item there is a clear preference towards a relatively close alignment between the orbital plane and the circumstellar disks.  Although some best-fit orbits are coplanar with the circumstellar disks within measurement uncertainties, as a whole there is a systematic tilt of up to typically $\sim$25\degr\ between the orbital plane of the vast majority of best-fit orbits and the two circumstellar disks

\end{itemize}

\noindent  By requiring that the radii of the two circumstellar disks should not exceed their predicted tidally-truncated radii, the best-fit orbits are restricted to primarily those having low eccentricities ($e \lesssim 0.2$).  The systematic tilt between these best-fit orbits and the circumstellar disks are now confined entirely to within $\sim$25\degr.  Thus, {\it the orbital plane is either coplanar or, more probably, tilted by up to a few tens of degrees from the two circumstellar disks.}  

We made a map of the inner region of the flattened dust envelope around L1551\,IRS\,5 based on data taken with the SMA by \citet{Chou2014}.  By fitting a Gaussian function to the image, we found that:

\begin{itemize}

\item the inner region of the flattened envelope has an inclination of  $54\degr.1 \pm 1\degr.0$ and a position angle for its major axis $156\degr.6 \pm 1\degr.3$, closely comparable with the corresponding parameters for the two circumstellar disks

\end{itemize}

\noindent We confirmed that:

\begin{itemize}

\item the protostars are orbiting each other in a clockwise fashion, similar to the clockwise rotation of their surrounding flattened envelope

\end{itemize}

Thus, {\it the two protostars have circumstellar disks that are closely parallel with their surrounding envelope, and are orbiting each other in the same direction as the envelope rotation.}  {This close alignment suggests that the angular momentum of material that gave rise to the protostellar system and that comprising its parental core are aligned.}

Fragmentation driven by {local (small-scale)} turbulence that conspires to generate local angular momentum can produce a binary system having relatively well aligned circumstellar disks and orbital planes.  In such cases, however, the binary system would not be expected to share a close geometrical or kinematic relationship with its surrounding envelope (remnant parental core).  On the other hand, the close geometrical and kinematic relationships seen in L1551\,IRS\,5 are expected for binary systems that form through rotationally-driven fragmentation of their flattened parental core.  In support of this argument, both the measured rotational velocity and the high degree of flattening exhibited by the envelope around L1551\,IRS\,5 suggest that its parental core possessed considerable angular momentum.  The probable misalignment between the circumstellar disks/flattened envelope and the orbital plane of L1551\,IRS\,5 may reflect the formation of the binary fragments in different planes of their flattened parental core sharing the same rotational axis.  The likely low eccentricity for the orbit of L1551\,IRS\,5 may have been imprinted at birth when the more massive envelope acted to circularize the binary orbit.  

Finally, we reason that L1551\,IRS\,5 will likely evolve into a solar-mass binary system that has roughly equal companion masses, quite closely coplanar circumstellar disks and orbital planes (especially if tidal interactions between one protostar and the circumstellar disk of the other protostar bring their circumstellar disks further into alignment with the orbital plane), and hence presumably also closely aligned stellar spin and orbital axes.  Unless its orbital parameters are dramatically altered during its subsequent evolution, L1551\,IRS\,5 will probably have a circular or low eccentricity orbit with a semimajor axis of many tens to perhaps a few hundred AU, and a corresponding orbital period of many hundreds to perhaps a few thousand years.  In this manner, L1551\,IRS\,5 will grow to resemble the vast majority of binary systems on the pre-main-sequence and main-sequence having comparable masses and orbital separations; such systems, which span a broad range of orbital eccentricities, exhibit preferentially aligned circumstellar disks and orbital planes or stellar spin and orbital axes.

\acknowledgments
J. Lim acknowledges support from the Research Grants Council of Hong Kong through grant HKU\,703512P.  T. Hanawa acknowledges support from the Japan Society for the Promotion of Science (JSPS) Grants-in-Aid for Scientific Research (KAKENHI) through Grant Number JP15K05017, and T. Matsumoto from the JSPS KAKENHI Grant Number 26400233.  S.Takakuwa acknowledges support from the Ministry of Science and Technology (MOST) of Taiwan through grant 102-2119-M-001-012-MY3.



{\it Facilities:} \facility{Jansky Very Large Array}.

\clearpage

\clearpage

\clearpage

\clearpage

\clearpage

\clearpage

\clearpage

\begin{figure}
\epsscale{0.9}
\center
\plotone{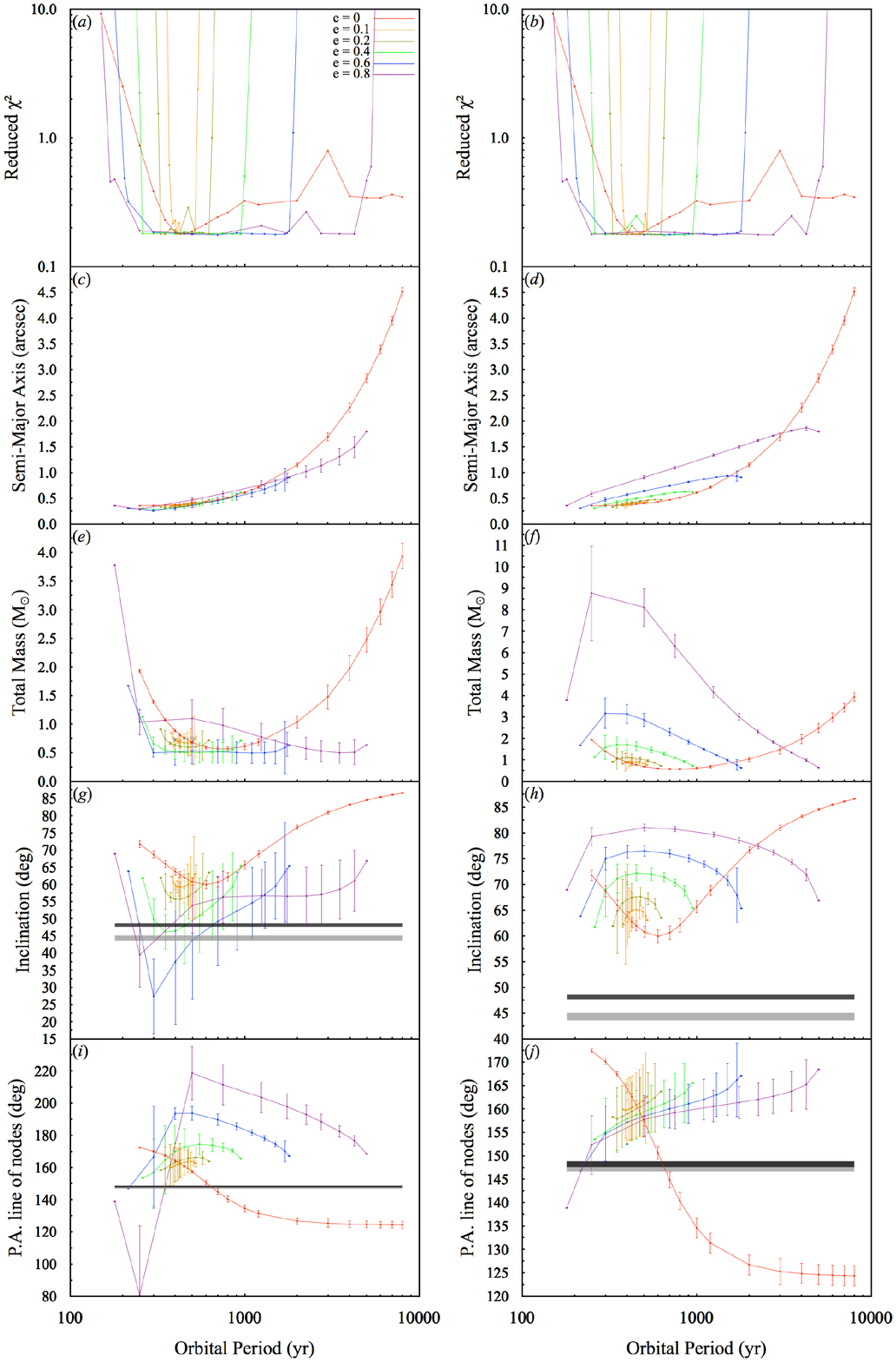}
\end{figure}
\begin{figure}
\renewcommand{\baselinestretch}{1.0}
\caption{\small Best-fit orbits to the relative proper motion shown in Fig.\,\ref{proper motion} computed at selected orbital periods (data points with error bars indicating $\pm 1\sigma$ uncertainties), showing $(a)$--$(b)$ the reduced $\chi^2$ of the fit, $(c)$--$(d)$ semi-major axis of the orbit, $(e)$--$(f)$ total mass of the binary system, $(g)$--$(h)$ inclination of the orbit, and $(i)$--$(f)$ position angle of the orbital line of nodes.  {Different colors correspond to the different orbital eccentricities considered as indicated in the top right corner of panel $(a)$}.  Curves correspond to linear interpolations between the data points.  The two columns correspond to the two families of solutions (left column family A, right column family B; see text) that provide acceptable fits (i.e., having reduced $\chi^2 \leq 1$) .  The horizontal lines in panels $(g)$--$(h)$ indicate the inclinations and those in panels $(i)$--$(j)$ the position angles of the major axes of the circumstellar disks for the N (black) and S (gray) sources, and have vertical widths corresponding to the $\pm 1\sigma$ uncertainties of these parameters.}
\label{Best-Fit Orbits}
\end{figure}
\clearpage

\clearpage

\begin{figure}
\epsscale{0.9}
\center
\plotone{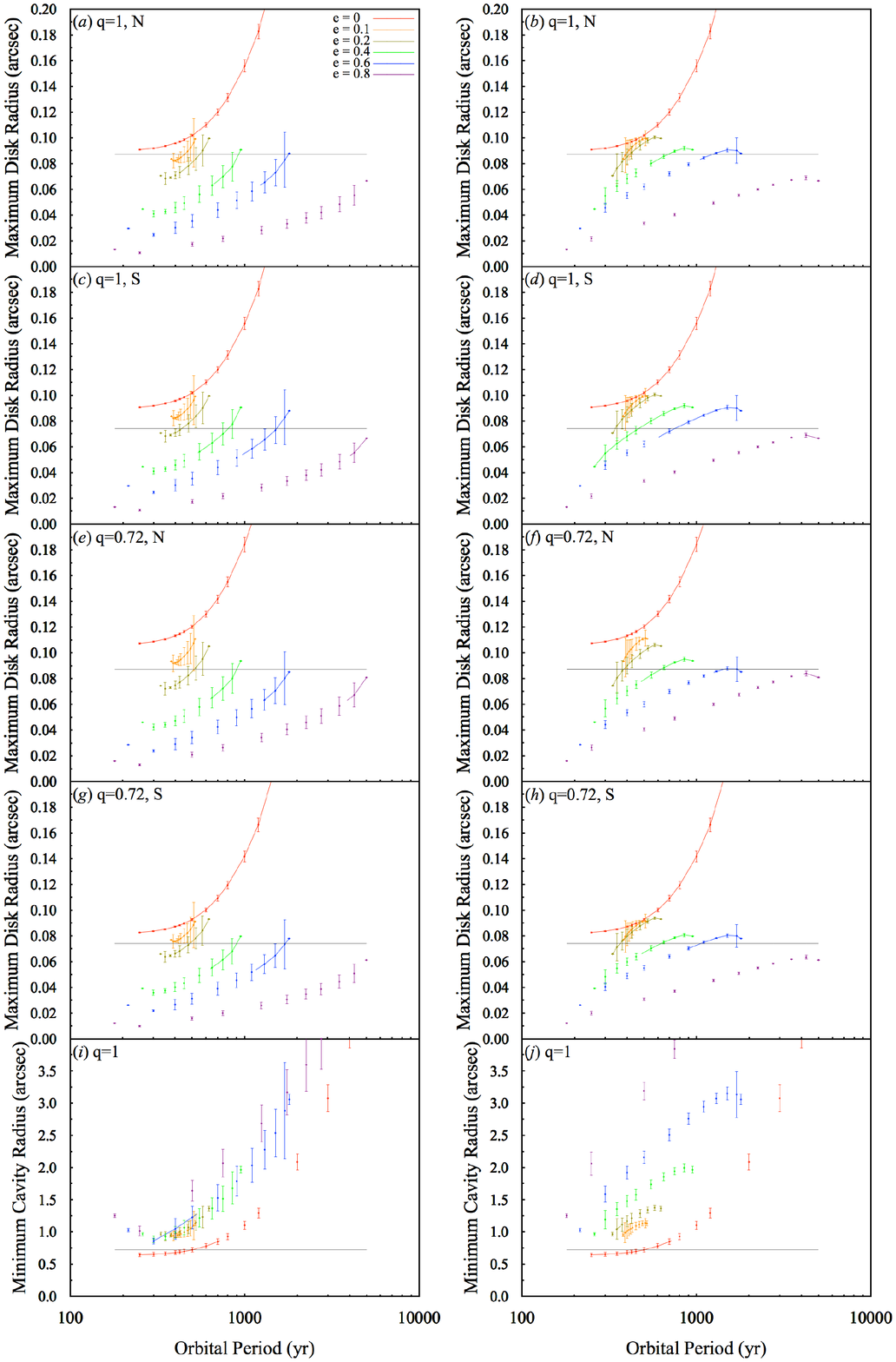}
\end{figure}
\begin{figure}
\renewcommand{\baselinestretch}{0.85}
\caption{\small Same best-fit orbits as in Fig.\,\ref{Best-Fit Orbits}, showing the maximum predicted extent for the circumstellar disk of the $(a)$--$(b)$ N source and $(c)$--$(d)$ S source as imposed by tidal truncation for a binary mass ratio $q=1$; same for the $(e)$--$(f)$ N source and $(g)$--$(h)$ S source but for a binary mass ratio $q=0.72$; and $(i)$--$(j)$ the minimum radius of a central gap in the envelope imposed by tidal truncation.  The break radius for the circumstellar disk of the N source is indicated by the horizontal line in $(a)$--$(b)$ and $(e)$--$(f)$, and that of the S source by the horizontal line in $(c)$--$(d)$ and $(g)$--$(h)$.  The curves in $(a)$--$(h)$ indicate the best-fit orbits for which the predicted tidally-truncated radius is, to within an uncertainty of $3\sigma$, no smaller than the break radius.  The FWHM of the Gaussian function fitted to the dust envelope shown in Fig.\,\ref{Pseudodisk} is indicated by the horizontal line in $(i)$--$(j)$.  The curves in these panels correspond to best-fit orbits in which the predicted minimum size of a central cavity in the envelope is, to within an uncertainty of $3\sigma$, no larger than the FWHM of the dust envelope.}
\label{Orbital Constraints}
\end{figure}
\clearpage

\clearpage

\begin{deluxetable}{ccccc}
\vspace{0cm}
\tabletypesize{\normalsize}
\tablecolumns{5}
\tablewidth{0pc}
\tablecaption{Map Parameters \label{Map Parameters}}
\tablehead{
\colhead{Map} & \colhead{\hspace{2.0cm} Synthesized Beam} & & & \colhead{rms} \\
\colhead{Weighting} & \colhead{\hspace{-3.5cm} Major Axis} & \colhead{\hspace{-5cm} Minor Axis} & \colhead{\hspace{-1.5cm} Position Angle} & \colhead{Noise} \\
\colhead{} & \colhead{\hspace{-3.5cm} (mas)} & \colhead{\hspace{-5cm} (mas)} & \colhead{\hspace{-2cm} (deg)} & \colhead{($\mu$Jy)} }
\startdata
Natural & \hspace{-3.5cm} 55.62 & \hspace{-5cm} 52.66 & \hspace{-1.5cm} $-1.65$ & 11.5 \\
$\rm Robust = -1.0$ & \hspace{-3.5cm} 33.40 & \hspace{-5cm} 31.74 & \hspace{-1.5cm} $-24.12$ & 35.4 \\
\enddata
\end{deluxetable}

\begin{deluxetable}{cccccccccc}
\vspace{0cm}
\tabletypesize{\scriptsize}
\tablecolumns{10}
\tablewidth{0pc}
\tablecaption{Parameters of Gaussian fits\label{Fitting Parameters}}
\tablehead{
\colhead{Map} & \colhead{Feature} & \colhead{Right} & \colhead{Declination}  & \colhead{Peak} & \colhead{Integrated} & \colhead{Major} & \colhead{Minor} & \colhead{Position} & \colhead{Inclination$\rm ^a$} \\
\colhead{Weighting} & & \colhead{Ascension} & \colhead{}  & \colhead{Intensity} & \colhead{Intensity} & \colhead{Axis} & \colhead{Axis} & \colhead{Angle} & \colhead{} \\
\colhead{} & & \colhead{(J2000)} & \colhead{(J2000)}  & \colhead{(mJy)} & \colhead{(mJy)} & \colhead{(mas)} & \colhead{(mas)} & \colhead{(deg)} & \colhead{(deg)}}
\startdata
\multicolumn{10}{c}{\footnotesize N source} \\
\tableline
\vspace{-0.2cm}  & Circumstellar & 04:31:34.15781(2) & 18:08:04.7933(4) & 1.45(1) & 6.43(6) & 134(1) & 97(1) & 149.2(9) &  \\
\vspace{-0.4cm}  & Dust Disk & & & & & \hspace{-1.9cm} deconvolved: 122(1) & 81(1) & 148.2(7) & $48.1 \pm 0.5$ \\ 
\vspace{-0.1cm} Natural & & & & & & & & & \\
\vspace{-0.2cm}   & Ionized & 04:31:34.15666(7) & 18:08:04.790(1) & 0.30(1) & 0.35(2) & 64(2) & 53(2) & 34.7(8.6) & \\
\vspace{-0.1cm} & Jet & & & & & \hspace{-1.7cm} deconvolved: 33(29) & \nodata & 42.3(8.9) & \\ 
\vspace{0.2cm} & & & & & &  \hspace{-2.7cm} synthesized beam: 55.6 & 52.7 & -1.6 & \\ 
\vspace{-0.2cm}  & Circumstellar & 04:31:34.15748(13) & 18:08:04.794(2) & 0.53(2) & 5.06(25) & 120(5) & 84(4) & 147.9(5.0) & \\
\vspace{-0.4cm}  & Dust Disk & & & & & \hspace{-1.9cm} deconvolved: 115(4) & 77(3) & 147.8(3.6) & $47.9 \pm 2.6$ \\ 
\vspace{-0.1cm} Robust & & & & & & & & & \\
\vspace{-0.2cm}  & Ionized & 04:31:34.15753(8) & 18:08:04.793(1) & 0.34(3) & 0.31(4) & 39(3) & 25(2) & 47.9(6.8) & \\
\vspace{-0.1cm} & Jet & & & & & \hspace{-1.9cm} deconvolved: 22(4) & \nodata & 49.7(4.8) & \\ 
\vspace{0.2cm} & & & & & & \hspace{-2.7cm} synthesized beam: 33.4 & 31.7 & -24.1 & \\
\tableline
\\
\multicolumn{10}{c}{\footnotesize S source} \\
\tableline
\vspace{-0.2cm}  & Circumstellar & 04:31:34.15983(2) & 18:08:04.4377(4) & 1.21(1) & 3.76(4) & 107(1) & 85(1) & 149.6(1.5) & \\
\vspace{-0.4cm}  & Dust Disk & & & & &  \hspace{-1.7cm} deconvolved: 92(1) & 66(1) & 147.7(1.2) & $44.3 \pm 0.8$ \\ 
\vspace{-0.1cm} Natural & & & & & & & & & \\
\vspace{-0.2cm}   & Ionized & 04:31:34.16516(11) & 18:08:04.476(2) & 0.30(1) & 1.29(5) & 122(4) & 102(4) & 25.6(7.6) & \\
\vspace{-0.1cm} & Jet & & & & & \hspace{-1.9cm} deconvolved: 109(3) & 87(3) & 27.2(6.2) & \\ 
\vspace{0.2cm} & & & & & & \hspace{-2.7cm} synthesized beam: 55.6 & 52.7 & -1.6 & \\ 
\vspace{-0.2cm}  & Circumstellar & 04:31:34.1600(1) & 18:08:04.438(2) & 0.54(2) & 3.16(17) & 88(4) & 71(3) & 150.3(8.5) & \\
\vspace{-0.4cm}  & Dust Disk & & & & &  \hspace{-1.9cm} deconvolved: 81(3) & 63(3) & 150.1(6.4) & $38.8 \pm 4.3$ \\ 
\vspace{-0.1cm} Robust & & & & & & & & & \\
\vspace{-0.2cm}   & Ionized & 04:31:34.1651(3) & 18:08:04.488(6) & 0.13(3) & 0.30(8) & 79(16) & 32(6) & 39.9(7.9) & \\
\vspace{-0.1cm} & Jet & & & & & \hspace{-2.0cm} deconvolved: 73(6) & \nodata & 40.3(5.7) & \\ 
\vspace{0.2cm} & & & & & & \hspace{-2.7cm} synthesized beam: 33.4 & 31.7 & -24.1 & \\
\enddata
\tablenotetext{a}{Inclination = arccos(Minor Axis/Major Axis)}
\tablecomments{Numbers in brackets indicates the root-mean-square uncertainty in the corresponding final digit or digits.}
\end{deluxetable}
\clearpage

\begin{deluxetable}{cccccccc}
\tabletypesize{\normalsize}
\tablecolumns{8}
\tablewidth{0pc}
\tablecaption{Parameters of NUKER fits \label{Parameters NUKER fits}}
\tablehead{
\colhead{Map} & \colhead{Inclination} & \colhead{Position Angle} & \colhead{\hspace{0.2cm}$\gamma$} & \colhead{\hspace{0.2cm}$\beta$} & \colhead{\hspace{0.2cm}$\alpha$} & \colhead{\hspace{1.0cm} $r_b$} \\
\colhead{Weighting} & & \colhead{Major Axis} & \colhead{} & \colhead{} & \colhead{} & \colhead{}  \\
\colhead{}  & \colhead{(deg)} & \colhead{(deg)} & \colhead{} & \colhead{} & \colhead{} & \colhead{\hspace{-0.2cm}(mas)} & \colhead{\hspace{-0.5cm} (AU)} }
\startdata
\multicolumn{8}{c}{N source} \\
\tableline
Natural & $48.1$ & $148.2$ & $\hspace{0.2cm}0.60$ & $\hspace{0.2cm}8.7$ & $\hspace{0.2cm}37.5$ & $87.4$ & $12.2$ \\
Robust & $47.9$ & $147.8$ & $\hspace{0.2cm}0.54$ & $\hspace{0.2cm}11.5$ & $\hspace{0.2cm}40.0$ & $81.4$ & $11.4$ \\
\tableline
\\
\multicolumn{8}{c}{S source} \\
\tableline
Natural & $44.3$ & $147.7$ & $0.81$ & $10.3$ & $\hspace{0.0cm}37.5$ & $73.9$ & $10.3$ \\
\enddata
\tablecomments{Inclinations and position angles are fixed at their formal values derived from the deconvolved parameters of the circumstellar disks based on the two-component two-dimensional Gaussian function fitted to each source as listed in Table\,\ref{Fitting Parameters}.}
\end{deluxetable}
\clearpage

\begin{deluxetable}{cccc}
\tabletypesize{\normalsize}
\tablecolumns{4}
\tablewidth{0pc}
\tablecaption{Parameters of Flattened Envelope \label{Parameters Pseudodisk}}
\tablehead{
\colhead{Major Axis} & \colhead{Minor Axis} & \colhead{Position Angle} & \colhead{Inclination} \\
\colhead{(arcsec)} & \colhead{(arcsec)} & \colhead{(deg)} & \colhead{(deg)}}
\startdata
$1.44 \pm 0.02$ & $0.85 \pm 0.02$ & $156.6 \pm 1.3$ & $54.1 \pm 1.0$ \\
\enddata
\tablecomments{Size of major and minor axes correspond to deconvolved FWHM of Gaussian function fitted to the dust emission of Figure\,\ref{Pseudodisk}.  Inclination derived from the values listed for the major and minor axes.}
\end{deluxetable}
\clearpage

\end{document}